\documentclass[a4paper,11pt]{article}
\usepackage{jheppub}
\usepackage{mathrsfs}
\usepackage{amsfonts}
\usepackage{setspace}
\usepackage{cellspace}
\usepackage{amsmath,amssymb,bm}
\usepackage[colorlinks=true,linkcolor=blue]{hyperref}
\usepackage{xcolor}
\usepackage{epsfig}
\usepackage{slashed}
\usepackage{caption}
\usepackage{hhline,multirow,tabularx}  % for nicer tables
\usepackage{dcolumn}    % align table columns on decimal point
\usepackage{url}        % for URL addresses
\usepackage{braket} 
\definecolor{cyan}{rgb}{0.0, 1.0, 1.0}
\definecolor{applegreen}{rgb}{0.55, 0.71, 0.0}
\definecolor{arylideyellow}{rgb}{0.91, 0.84, 0.42}
\definecolor{bananayellow}{rgb}{1.0, 0.88, 0.21}
\definecolor{burlywood}{rgb}{0.87, 0.72, 0.53}
\definecolor{buff}{rgb}{0.94, 0.86, 0.51}
\definecolor{blond}{rgb}{0.98, 0.94, 0.75}
\definecolor{bisque}{rgb}{1.0, 0.89, 0.77}
\definecolor{bananamania}{rgb}{0.98, 0.91, 0.71}
\definecolor{apricot}{rgb}{0.98, 0.81, 0.69}
\definecolor{almond}{rgb}{0.94, 0.87, 0.8}

\title{Bjorken and threshold asymptotics of a space-like structure function in the 2D $U(N)$ Gross-Neveu model}
\author[a]{Yizhuang Liu}
\affiliation[a]{Institute of Theoretical Physics,
Jagiellonian University, 30-348 Kraków, Poland}
\emailAdd{yizhuang.liu@uj.edu.pl}
\abstract {In this work, we investigate a coordinate space structure function ${\cal E}(z^2m^2,\lambda)$ in the 2D $U(N)$ Gross-Neveu model to the next-to-leading order in the large-$N$ expansion. We analytically perform the twist expansion in the Bjorken limit through double Mellin representations. Hard and non-perturbative scaling functions are naturally generated in their Borel representations with detailed enumerations and explicit expressions provided to all powers. The renormalon cancellation at $t=n$ between the hard functions at powers $p$ and the non-perturbative functions at powers $p+n$ are explicitly verified, and the issue of ``scale-dependency'' of the perturbative and non-perturbative functions is explained naturally. Simple expressions for the leading power non-perturbative functions are also provided both in the coordinate space and the momentum-fraction space ($0<\alpha<1$) with ``zero-mode-type'' subtractions at $\alpha=0$ discussed in detail. In addition to the Bjorken limit, we also perform the threshold expansion of the structure function up to the next-to-next-to-leading threshold power exactly and investigate the resurgence relation between threshold and ``Regge'' asymptotics. We also prove that the twist expansion is absolutely convergent for any $0<z^2<\infty$ and any $ \lambda \in iR_{\ge 0}$. }
\date{\today}

\begin{document}
\maketitle
\flushbottom
\section{Introduction}
In this paper, we consider a specific structure function in the 2D $U(N)$ Gross Neveu model with non-perturbative quark mass $m$, to next-to-leading order in the $\frac{1}{N}$ expansion. The purpose is fourfold:
\begin{enumerate}
    \item To provide an explicit derivation of the specific structure of the marginal asymptotics in the Bjorken limit $z^2\rightarrow 0$ at fixed $\lambda$. In particular, to calculate analytically the perturbative and non-perturbative functions appearing in the expansion.     
 \item To demonstrate explicitly the pattern of renormalon cancellation~\cite{Shifman:1978bx,David:1983gz,David:1985xj,Mueller:1984vh,Ji:1994md,Beneke:1998ui,Beneke:1998eq,Braun:2004bu,  Shifman:2013uka,Dunne:2015eoa} to all powers, and investigate other important quantitative properties of the expansion such as the convergence property. 
    \item To perform systematically the threshold expansion in $\frac{1}{\lambda}$, demonstrate its convergence/divergence property, verify the resurgence relation to the ``Regge'' asymptotics~\cite{Basso:2006nk} and investigate its relation to the twist expansion.
    \item To demonstrate through the concrete calculations a number of essential features of perturbative expansions/power expansions in QCD-like theories that are confusing for beginners.

\end{enumerate}
Notice that although we performed the $\frac{1}{N}$ expansion,  at any given order $\frac{1}{N^k}, k\ge 1$ , there are infinitely many powers in $z^2m^2$. Moreover, at any given power, there 
are infinitely many non-perturbative matrix elements of fixed-twist operators, each attached to an OPE coefficient (also called hard coefficient) with infinitely many perturbative orders. At least to each order in $\frac{1}{N}$, the matrix elements and OPE coefficients at any given power can be further organized into a finite number of $\lambda$-dependent non-perturbative scaling functions and hard functions. The hard functions, as well the non-perturbative functions, can suffer from scale/scheme ambiguities that have to be canceled within each power at the logarithmic level, as well as the renormalon singularities that are finally canceled between  perturbative and non-perturbative contributions across different powers.  

To our knowledge, this type of expansions in the {\it Bjorken limit} to the level that allows to see the coexistence of both non-perturbative functions as well as perturbative hard kernels with manifest renormalon-cancellation pattern,  has never been analytically worked out in 2D QFTs with marginal UV limits, even at the level of large $N$ expansion (expansions in the Euclidean high-virtuality limit $-p^2\rightarrow +\infty$ for vacuum correlators have been worked out many years ago for the $O(N)$ non-linear sigma model in~\cite{Beneke:1998eq} and for the large-$N$ Gross Neveu recently in~\cite{Marino:2024uco}). On the other hand, the Bjorken limit in the presence of marginality is conceptually as well as phenomenologically important, since it is rooted in one of the benchmark processes of QCD, the space-like DIS, in the context of which such expansions are commonly called ``twist expansion'' or ``collinear factorization'' and the non-perturbative scaling functions appearing in the expansion are commonly called ``parton distribution functions (PDFs)''. 

Since there are no known approximations to QCD that simultaneously work at the scale of $\Lambda_{\rm QCD}$ and match with the PQCD in UV, the specific structure of the QCD Bjorken limit commonly called ``collinear factorization'', strictly speaking, has no exact derivation up to now. As such,  it is valuable to find exactly workable examples in which the marginal asymptotics in the Bjorken limit in terms of perturbative hard functions and non-perturbative ``PDFs'' can be explicitly and exactly derived with full control of the non-perturbative functions as well as high-order perturbative growths, a task that has not been done before in the literature.

In the rest of the work, we will try to fill the gap in the literature by analytically performing the ``twist-expansion'' in the Bjorken limit for a {\it coordinates space} structure function in the 2D large-$N$ Gross-Neveu model at order $\frac{1}{N}$. We hope this work can serve as a simple yet non-trivial demonstration-of-principle example for the ``twist-expansion'', as well as for many interesting mathematical features and subtleties related to various kinematic limits of ``structure-function-type'' quantities as well as the perturbative expansion/power expansion themselves in marginal theories.

As a reminder~\cite{Coleman:1985rnk}, the Lagrangian of the theory is 
\begin{align}
{\cal L}=\bar \psi i\slashed{\partial}\psi-\sigma \bar\psi \psi-\frac{\sigma^2}{2g_0^2} \ .
\end{align}
The large-$N$ expansion is performed in the following way. The effective action for the $\sigma$ field reads
\begin{align}
i\int d^Dx {\cal L}(\sigma)=-\int d^Dx \frac{i(m+\sigma)^2}{2g_0^2}+N{\rm Tr} \ln (\slashed{p}-m-\sigma) \ .
\end{align}
Now, the condition that $\sigma_0=m$ is a saddle point implies the ``gap equation''
\begin{align}
\frac{1}{2g_0^2}=N\int \frac{d^Dp}{(2\pi)^D}\frac{i}{p^2-m^2} \ .
\end{align}
This leads to the renormalized coupling constant
\begin{align}
\frac{1}{2g^2(\mu_0)}=\frac{N}{4\pi}\ln \left(\frac{4\pi e^{-\gamma_E}\mu_0^2}{m^2}\right) \ .
\end{align}
By expanding the action in $\sigma$, the quadratic part of the $\sigma$ action is
\begin{align}
i\int dx^2 {\cal L}^{(2)}(\sigma)=-\frac{N}{2}\int \frac{d^2k}{(2\pi)^2} \sigma(k)S^{-1}(k^2)\sigma(-k) \ ,
\end{align}
where the propagator for the sigma field reads
\begin{align}
&S^{-1}(k^2)=(-k^2+4m^2)\int \frac{d^2p}{(2\pi)^2}\frac{1}{((p+k)^2-m^2)(p^2-m^2)} \nonumber \\
&=\frac{i}{4\pi}(-k^2+4m^2)\int_{0}^1 dx \frac{1}{-x(1-x)k^2+m^2}=\frac{i}{2\pi}A(-k^2)\ln \frac{A(-k^2)+1}{A(-k^2)-1}   \ ,
\end{align}
with
\begin{align}
A(-k^2)=\sqrt{1-\frac{4m^2}{k^2}} \ .
\end{align}
Notice that the above defines a unique analytic function in $k^2$ with a branch cut from $4m^2$ to $+\infty$ along the real axis when both the $\sqrt{1-\frac{4m^2}{k^2}}$ function and the logarithmic function are defined with the principal value prescription.   The above implies the effective coupling
\begin{align}
\frac{1}{g^2(-k^2)}=\frac{N}{2\pi}A(-k^2)\ln \frac{A(-k^2)+1}{A(-k^2)-1} \ .
\end{align}
Notice that $g^2(0)=\frac{\pi}{N}$ is finite. On the other hand, in the large $-k^2$ limit, one has
\begin{align}
\frac{1}{g^2(-k^2)} \rightarrow \frac{N}{2\pi}\ln \frac{-k^2}{m^2} \ .
\end{align}
This agrees with the $g(\mu^2)$ in the $\overline{MS}$ scheme, if one identifies the $\mu^2=4\pi e^{-\gamma_E}\mu_0^2$, as expected.

\section{Mellin-representations, twist expansion and non-perturbative functions}

Given the above, in this section we investigate the forward matrix element (which we call as a ``structure function'') 
\begin{align}\label{eq:def}
&\langle p,i|\bar \psi^i(x)\psi^i(0)|p,i\rangle-\langle p,i|\bar \psi^j(x)\psi^j(0)|p,i\rangle \equiv F(x,p) \equiv \bar u(p)u(p){\cal E}(z^2m^2,\lambda) \ ,
\end{align}
where $|p,i\rangle$ is an on-shell massive fermion state with flavor species $i$, $p^2=m^2$, and $\slashed{p}u(p)=mu(p)$. Here $i\ne j$ and no summation is assumed. Clearly, the subtraction removes the ``disconnected'' or ``sea'' contribution, leaving only the ``valence'' part of the structure function. Here $z^2=-x^2>0$ (spacelike separation) and $\lambda=-p\cdot x$ are the two Lorentz invariant combinations formed between $x$ and $p$. To next-to-leading order in $\frac{1}{N}$, there is essentially only one diagram (the ladder diagram), which we calculate below. Wave-function and mass renormalizations only contribute to  $z,\lambda$ independent numbers multiplying $e^{-i\lambda}\bar u(p) u(p)$, therefore will not be calculated. 

Notice that this type of correlator has been known since the very early days of QFTs,  since it is the versions in which  
the fermion fields were replaced by conserved currents that are commonly called ``structure functions'' and can be probed in DIS-like process~\cite{Itzykson:1980rh, Sterman:1993hfp, Collins:2011zzd}. On the other hand, most of the investigations before the 20th century are in the momentum space. The coordinate space versions in the QCD literature, to the author's knowledge, arise initially to investigate theoretical aspects of the twist expansion~\cite{Braun:2004bu} and attract considerable attention later in the context of lattice calculation~\cite{Ji:2013dva,Ji:2020ect,Cichy:2021ewm}. Due to this, the forward matrix element like the one in Eq.~(\ref{eq:def}) has received various names, such as the ``quasi-PDF''~\cite{Ji:2013dva} (normally refers to the {\it one-dimensional} Fourier transform along the spatial direction of the external momentum) and the ``pseudo-PDF''~\cite{Radyushkin:2017cyf} (normally refers to the coordinate space version in Eq.~(\ref{eq:def})). Compared with the momentum space versions, the coordinate space version is closer to the operator definitions for the ``parton distribution functions''~\cite{Collins:2011zzd}. In particular, the one-dimensional Fourier transform with respect to $\lambda$ is compactly supported in $[-1,1]$ and closely resembles the ``momentum-fraction'' of PDFs. 

There are several important limits that can be defined for the correlator in Eq.~(\ref{eq:def}):
\begin{enumerate}
    \item The Bjorken limit, defined as $z^2 \rightarrow 0$ at a fixed $\lambda$. Notice that in order to reach this limit, either the external state or the separation $x$ must be boosted to be close to the light front. As such, the various powers in $z^2m^2$ are classified naturally according to the ``twists'' of the operators (mass dimension minus spin) appearing in the OPE of $\bar \psi_i(x)\psi_i(0)$, since operators with Lorentz indices receive additional ``boost enhancements''. For the specific correlator in Eq.~(\ref{eq:def}), the twist starts from $1$. 
    \item The threshold limit, defined as $\lambda \rightarrow +i\infty$ at a fixed $z^2$ or on top of the small $z^2$ expansion. To explain this name, first notice that the Fourier conjugate $\alpha$ of $\lambda$ is supported in $[-1,1]$, implying analyticity in $\lambda$ in the whole complex plane. Then, if one chose $\lambda$ to be along the positive imaginary axis, then in the $-i\lambda \rightarrow +\infty$ limit,  one naturally probes the $\alpha \rightarrow 1$ limit of the distribution to maximize the exponential $e^{-i\alpha \lambda}$ ($-1<\alpha<1$). 
    \item The ``Regge'' limit, defined as $\alpha \rightarrow 0$ in the momentum-fraction space. For the valence correlator in Eq.~(\ref{eq:def}), at the order $\frac{1}{N}$, the $\alpha$ is actually supported in $[0,1]$. In this situation, the ``Regge'' limit can also be probed in the coordinate space as $\lambda \rightarrow -i\infty$. 
\end{enumerate}
In the rest of the work, we will perform a thorough investigation of the correlator in Eq.~(\ref{eq:def}) at the order $\frac{1}{N}$. Explicit expansions in the Bjorken limit will be performed to all powers. Expansions in the threshold limit will be performed explicitly to the next-to-next-to-leading power in $\frac{1}{\lambda}$ at generic $z^2m^2$. Although our method allows us to write down the general expressions at an arbitrary threshold power, we will only do so for a non-perturbative scaling function at the leading twist to investigate the ``resurgence relation'' to the ``Regge'' limit. 

\subsection{Mellin representations and overall structure of the twist expansion}
To proceed, notice first that
\begin{align}
\bar u (\slashed{p}+m-\slashed{k})(\slashed{p}+m-\slashed{k})u=(k^2+4m^2-4p\cdot k)\bar u u \ .
\end{align}
Given the above, one needs to consider the integral (in Minkowskian signature with $k^2$ understood as $k^2+i0$)
\begin{align}
&F(x,p)e^{i\lambda}=i\bar u u\int\frac{d^2k}{(2\pi^2)}g^2(-k^2)\frac{(-k^2-4m^2+4p\cdot k)e^{-ik\cdot x}}{((p-k)^2-m^2)^2} \ .
\end{align}
To calculate this integral,  one needs the crucial representation following the same spirit in Ref.~\cite{Beneke:1998eq}:
\begin{align}
g^2(-k^2)=\frac{2\pi}{N} \int_{0}^{\infty} dt \int_{c-i\infty}^{c+i\infty} \frac{ds}{2\pi i}\frac{\Gamma(1-2s)\Gamma(s+t)}{\Gamma(1-s+t)} \left(\frac{m^2}{-k^2}\right)^{-s} \ .
\end{align}
Now, the vertical line of the inverse Mellin transform has to be taken as
\begin{align}
-t<c \equiv {\rm Re}(s)<\frac{1}{2} \ ,
\end{align}
which always include the region $0<c<\frac{1}{2}$ which we use throughout the paper.  Now, the results read (as in~\cite{Liu:2023onm}, we analytically continue $\lambda$ to the positive imaginary axis and use the same letter $\lambda$ to denote the coordinate in the imaginary axis in all the functions except the ${\cal E}$ itself)
\begin{align}
{\cal E}(z^2m^2,s,i\lambda)=\frac{2\pi}{N}e^{\lambda}\bigg(-F_1(z^2m^2,s,\lambda)+F_2(z^2m^2,s,\lambda)-F_3(z^2m^2,s,\lambda)\bigg) \ , 
\end{align}
for which one has the parametric representations: 
\begin{align}
&F_1(z^2m^2,s,\lambda)=\frac{1}{4\pi \Gamma(-1-s)}\int_{0}^{\infty}\rho^{-s-1}d\rho\int_{0}^1 dx x(1-x)^{-2-s}e^{-\frac{z^2m^2}{4\rho}-\rho x^2-x\lambda} \ , \\ 
&F_2(z^2m^2,s,\lambda)=\frac{1}{\pi \Gamma(-s)}\int_{0}^{\infty}\rho^{-s}d\rho\int_{0}^1 dx x(1-x)^{-s}e^{-\frac{z^2m^2}{4\rho}-\rho x^2-x\lambda} \ , \\ 
&F_3(z^2m^2,s,\lambda)=\frac{\lambda}{2\pi \Gamma(-s)}\int_{0}^{\infty}\rho^{-s-1}d\rho\int_{0}^1 dx x(1-x)^{-s-1}e^{-\frac{z^2m^2}{4\rho}-\rho x^2-x\lambda} \ .
\end{align}
Combining the above with $\frac{\Gamma(1-2s)\Gamma(s+t)}{\Gamma(1-s+t)}$ and performing the $s,t$ integrals, one obtains the functions $F_1(z^2m^2,\lambda)$, $F_2(z^2m^2,\lambda)$ and $F_3(z^2m^2,\lambda)$. 

To proceed, we now introduce a second layer of Mellin-representations for the functions $F_k(z^2m^2,s,\lambda)$, $k=1,2,3$ with the variable $u$ with respect to $\frac{z^2m^2}{4}$. The second layer of Mellin-representation leads to the closed forms ($\,_1\tilde F_1$ is the regularized confluent Hypergeometric function)
\begin{align}
F_1(u,s,\lambda)=\frac{1}{4\pi}\Gamma(2+2s-2u)\Gamma(u)\Gamma(u-s)\,_1\tilde F_1(2+2s-2u,1+s-2u,-\lambda) \ , \\
{\rm Re}(s)<{\rm Re} (u)<{\rm Re}(s)+1 \ ,
\end{align}
\begin{align}
F_2(u,s,\lambda)=-\frac{s}{\pi}\Gamma(2s-2u)\Gamma(u)\Gamma(u-s+1)\,_1\tilde F_1(2s-2u,1+s-2u,-\lambda) \ , \\
0<{\rm Re} (u)<{\rm Re}(s) \ , 
\end{align}
and
\begin{align}
F_3(u,s,\lambda)=\frac{\lambda}{2\pi}\Gamma(2+2s-2u)\Gamma(u)\Gamma(u-s)\,_1\tilde F_1(2+2s-2u,2+s-2u,-\lambda) \ , \\
{\rm Re}(s)<{\rm Re} (u)<{\rm Re}(s)+1 \ .
\end{align}
Notice that $F_1(z^2m^2,s,\lambda)$,  $F_2(z^2m^2,s,\lambda)$ and $F_3(z^2m^2,s,\lambda)$ themselves are regular at $s=0$. In particular, $F_2$ vanishes at $s=0$. Given the above representations, one then shifts the contour of $u$ and $s$ to the left to perform the expansion. The crucial facts are:
\begin{enumerate}
    \item For $F_1$, there are two series of poles: $u=-l$ and $u=s-l$ with $l\ge 0$. From each term in the first series, one obtains a non-perturbative function $\hat q_1^{(l)}$ (given latter in Eq.~(\ref{eq:q1l})) at power $\left(\frac{z^2m^2}{4}\right)^l$ represented as double $t,s$ integrals. On the other hand, from each of the second series of poles, one obtains $\left(\frac{z^2m^2}{4}\right)^{-s+l}$, which still depends on $s$ (this leads to the ``mother-function'' ${\cal F}_1^{(l)}$ in Eq.~(\ref{eq:motherlF1})). One still needs to shift $s$ to the left to obtain the power expansion. From the $\Gamma(s+t)$, one obtains $\left(\frac{z^2m^2}{4}\right)^{t+l+p}$ with $p\ge 0$, which leads to hard coefficients $ {\cal H}_1^{l,p}$ in Eq.~(\ref{eq:H1lp}) at powers $\left(\frac{z^2m^2}{4}\right)^{l+p}$. On the other hand, the $\Gamma(s-l)$ also contains poles at $s=-p$ with $p\ge 0$, which generates the non-perturbative functions $q_1^{l,p}$ in Eq.~(\ref{eq:q1lp}) at powers $\left(\frac{z^2m^2}{4}\right)^{l+p}$.  The expansion of $F_3$ is almost identical to $F_1$ and leads to the set of non-perturbative and perturbative functions $\hat q_3^{(l)}$ in Eq.~(\ref{eq:q3l}), $q_3^{l,p}$ in Eq.~(\ref{eq:q3lp}) and ${\cal H}_3^{l,p}$ in Eq.~(\ref{eq:H3lp}) for $l\ge 0$, $p\ge 0$. 
    \item For $F_2$, the first series of poles  $u=-l$ with $l\ge 0$ remains the same, leading to a series of non-perturbative functions $q_2^{(l)}$ in Eq.~(\ref{eq:q2l}) at powers $\left(\frac{z^2m^2}{4}\right)^l$. On the other hand, the second series of poles corresponds to $u=s-l$ with $l\ge 1$. It will generate a series of hard coefficient functions ${\cal H}_2^{l,p}$ in Eq.~(\ref{eq:H2lp}) as well as another series of non-perturbative functions $q_2^{l,p}$ in Eq.~(\ref{eq:q2lp}) at powers $\left(\frac{z^2m^2}{4}\right)^{l+p}$ with $p\ge 0$.
   \item Notice that the hard functions as well as non-perturbative functions generated by picking up poles from $\Gamma(t+s)\Gamma(s-l)$ are all represented as singular integrals in $t$ with poles at $t=n$, $n\ge 0$. Clearly, these poles occur at values of $t$ such that poles from the two gamma functions collide with each other. As such, all the $t$ integral poles must cancel between hard and non-perturbative contributions. In fact, we will show in Sec.~\ref{sec:generaltwist} that for each $l$, singularities at $t=n$ cancel for each $p\ge 0$, between ${\cal H}_{k}^{l,p}$ and  $q^{l,p+n}_{k}$ for $k=1,2,3$. 
   At $t=0$, this is the UV divergence cancellation at the logarithmic level. At $t>0$, this is usually called ``renormalon cancellation''. 
   \item Notice also that the non-perturbative functions $\hat q_1^{(l)}$ and $\hat q_3^{(l)}$ obtained by picking up poles at $u=-l$, although represented as Borel integrals, contain no singularity in the positive real axis $t\ge 0$, can still suffer $\frac{1}{t}$ tail as $t\rightarrow \infty$. These singularities always cancel with the ones in the large $t$ limit of the functions ${q}_1^{l,0}$ obtained by first picking up $u=s-l$, then picking up $s=0$. This is another reflection of the UV-IR cancellation at the logarithmic level that introduces scheme and scale dependency to hard and ``collinear'' functions. 
\item Given the $t=0$ singularity or the large $t$-tail cancellation, one must introduce additional scheme/scale choices to make sense of the hard and non-perturbative functions after Borel integrals, {\it in addition to} the standard $\pm i\epsilon$ or principal value prescriptions for the renormalon singularities at $t=n$ with $n\ge 1$. This leads to {\it scale dependent} perturbative and non-perturbative functions introduced latter in Sec.~\ref{sec:scalescheme} and~\ref{sec:generaltwist}
\end{enumerate}
In the rest of this subsection, for illustration purpose we present explicit results up to the next-to-leading power in $z^2m^2$. Details of the expansion to all powers are postponed to Sec.~\ref{sec:generaltwist}. The general structure of the expansion is similar to the ones for the vacuum two-point function investigated in~\cite{Beneke:1998eq}.
\subsection{Leading power and A-series of NLP}
The leading power contains : $u=0$ from $F_1$, $F_2$ and $F_3$; $u=s$ and then $s=0$, $s=-t$ from $F_1$ and $F_3$. The A series of NLP contains $u=s$ from $F_1$ and $F_3$ and then $s=-1$, $s=-t-1$. Notice that here for notation simplicity we denote $\hat q_1\equiv \hat q_1^{(0)}$, $\hat q_3 \equiv \hat q_3^{(0)}$ and $q_2 \equiv q_2^{(0)}$ in terms of the general notations in Sec.~\ref{sec:generaltwist}. We also have ${\cal H}^{(0)}\equiv {\cal H}_1^{0,0}+q_1^{0,0}+{\cal H}_3^{0,0}+q_3^{0,0}$ and ${\cal H}_A^{(1)} \equiv {\cal H}_1^{0,1}+{\cal H}_3^{0,1}$, $q_A^{(1)} \equiv q_1^{0,1}+q_3^{0,1}$. The large $t$ singularities cancel within leading power. The $t=1$ renormalon singularity cancels between LP and A series of NLP. 

First, the LP ``PDF'' reads\footnote{To include the wave-function and mass renormalization  contributions, simply add them to $q^{(0)}$. All other functions are not affected. }
\begin{align}\label{eq:q0}
q^{(0)}(\lambda)=\int_{0}^{\infty} dt \bigg(-\hat q_1(t,\lambda)-\hat q_3(t,\lambda)+q_2(t,\lambda)\bigg)\ ,
\end{align}
with
\begin{align}
&\hat q_1(t,\lambda)=\frac{1}{4\pi}\int_{c-i\infty}^{c+i\infty}\frac{ds}{2\pi i} \frac{\Gamma(1-2s)\Gamma(s+t)\Gamma(-s)\Gamma(2+2s)}{\Gamma(1-s+t)}\,_1\tilde F_1(2+2s,1+s,-\lambda) \ , \label{eq: q1}\\ 
&\hat q_3(t,\lambda)=\frac{\lambda}{2\pi}\int_{c-i\infty}^{c+i\infty}\frac{ds}{2\pi i} \frac{\Gamma(1-2s)\Gamma(s+t)\Gamma(-s)\Gamma(2+2s)}{\Gamma(1-s+t)}\,_1\tilde F_1(2+2s,2+s,-\lambda)\ , \label{eq:q3} \\ 
&q_2(t,\lambda)=-\frac{1}{\pi}\int_{c-i\infty}^{c+i\infty}\frac{ds}{2\pi i} \frac{s\Gamma(1-2s)\Gamma(s+t)\Gamma(-s+1)\Gamma(2s)}{\Gamma(1-s+t)}\,_1\tilde F_1(2s,1+s,-\lambda) \label{eq:q2}\ .
\end{align}
Both $\hat q_1(t,\lambda)$, $\hat q_3(t,\lambda)$ and $q_2(t,\lambda)$ are finite functions in $t$, but $\hat q_1(t,\lambda)$ and $\hat q_3(t,\lambda)$ decay at large $t$ only at the $\frac{1}{t}$ speed (the $\frac{1}{t}$ tail can be read from the pole at $s=0$), namely
\begin{align}
\hat q_1(t,\lambda)|_{t\rightarrow \infty}\rightarrow-\frac{1}{4\pi} \frac{1}{t}\,_1\tilde F_1(2,1,-\lambda)+{\cal O}\left(\frac{1}{t^2}\right) \ , \\ 
\hat q_3(t,\lambda)|_{t\rightarrow \infty}\rightarrow-\frac{\lambda}{2\pi} \frac{1}{t}\,_1\tilde F_1(2,2,-\lambda)+{\cal O}\left(\frac{1}{t^2}\right) \ ,
\end{align}
On the other hand, the hard kernel at LP is generated from 
\begin{align}
{\cal F}_A(z^2m^2,t,\lambda)=\frac{1}{4\pi}&\int_{c-i\infty}^{c+i\infty}\frac{ds}{2\pi i}  \left(\frac{z^2m^2}{4}\right)^{-s} \frac{\Gamma(1-2s)\Gamma(s+t)\Gamma(s)\Gamma(2)}{\Gamma(1-s+t)}\nonumber \\ 
&\times \bigg(\,_1\tilde F_1(2,1-s,-\lambda)+2\lambda\,_1\tilde F_1(2,2-s,-\lambda)\bigg)\ .
\end{align}
Picking up the poles at $s=0$ and $s=-t$, one has
\begin{align}\label{eq:h0}
{\cal H}^{(0)}(t,\alpha(z),\lambda)=&\frac{1}{4\pi}\bigg(\frac{1}{t}\,_1\tilde F_1(2,1,-\lambda)+\left(\frac{z^2m^2}{4}\right)^{t}\Gamma(-t)\,_1\tilde F_1(2,1+t,-\lambda)\bigg) \nonumber \\ 
+&\frac{\lambda}{2\pi}\bigg(\frac{1}{t}\,_1\tilde F_1(2,2,-\lambda)+\left(\frac{z^2m^2}{4}\right)^{t}\Gamma(-t)\,_1\tilde F_1(2,2+t,-\lambda)\bigg) \ .
\end{align}
Here we introduced the running coupling constant (it corresponds to $\frac{N g^2}{2\pi}$)
\begin{align}
\alpha(z) \equiv \frac{1}{\ln \frac{1}{z^2m^2}} \ .
\end{align}
Clearly, the large $t$ asymptotics has the $\frac{1}{t}$ tail, which cancels the ones for $\hat q_1(t,\lambda)$ and $\hat q_3(t,\lambda)$. One can show that up to treatments of the $\frac{1}{t}$ singularity at $t=0$, the first line in ${\cal H}^{(0)}(t,\alpha(z),\lambda)$ simply relates to the Borel transform of the perturbative coefficient function calculated through the same structure function for an on-shell massless quark and with a massless bubble chain. On the other hand, the second line can not be reproduced naively from massless partonic calculations with only one on-shell quark in the external state. This ``puzzel'' can be resolved by noticing that for our purpose, at leading power there are two series of operators in the OPE ($\{\mu_1\mu_2...\mu_n\}$ means the symmetric-traceless projection):
\begin{align}\label{eq:leadingtiwst}
\bar \psi_i(x)\psi_i(0)=&\sum_{n=0}^{\infty}\frac{1}{n!} {\cal H}_n(\alpha(z))x_{\mu_1}...x_{\mu_n}\bar\psi_i \overleftrightarrow{\partial}^{\{\mu_1}...\overleftrightarrow{\partial}^{\mu_n\}}\psi_i \nonumber \\ 
+&\sum_{n=0}^{\infty}\frac{1}{n!}\tilde {\cal H}_n(\alpha(z))x_{\mu_1}x_{\mu_2}...x_{\mu_{n+1}}\bar\psi_i \gamma^{\{\mu_1} \overleftrightarrow{\partial}^{\mu_2}...\overleftrightarrow{\partial}^{\mu_{n+1}\}}\psi_i \bar \psi\psi+... \ . 
\end{align}
Due to the presence of the VEV $g_0^2\langle \bar \psi \psi\rangle=-m+{\cal O}\left(\frac{m}{N}\right)$, contributions from the second series of operators are non-vanishing at order $\frac{1}{N}$ . One can show that the second line of Eq.~(\ref{eq:leadingtiwst}) exactly explains the second line of Eq.~(\ref{eq:h0}). 

One then moves to the series A of NLP. By picking up poles at $s=-1$ and $s=-t-1$, one has (notice that they are multiplied with the overall factor $\frac{z^2m^2}{4}$)
\begin{align}
&q^{(1)}_A(t,\lambda)=-\frac{1}{4\pi}\frac{\Gamma(3)\Gamma(t-1)}{\Gamma(2+t)}\,_1\tilde F_1(2,2,-\lambda) -\frac{\lambda}{2\pi}\frac{\Gamma(3)\Gamma(t-1)}{\Gamma(2+t)}\,_1\tilde F_1(2,3,-\lambda)\ . \label{eq:qA}
\end{align}
and
\begin{align}
&{\cal H}^{(1)}_A(t,\alpha(z),\lambda)\nonumber \\ &=-\frac{1}{4\pi}\left(\frac{z^2m^2}{4}\right)^{t}\frac{\Gamma(3+2t)\Gamma(-t-1)}{\Gamma(2+2t)}\bigg(\,_1\tilde F_1(2,2+t,-\lambda) +2\lambda \,_1\tilde F_1(2,3+t,-\lambda) \bigg) \label{eq:ha}  \ . 
\end{align}
One can see now that the $t=1$ renormalon pole cancels between ${\cal H}^{0}(t, \alpha(z),\lambda)$ and $\frac{z^2m^2}{4}q^{(1)}_A(t,\lambda)$, while the $t=0$ pole cancels between the two contributions in the A-series NLP.

\subsection{B-Series and C-series of NLP}
The B-series of NLP contains: $u=-1$ from $F_1$ and $F_3$; $u=s-1$ and then $s=0$, $s=-t$ from $F_1$ and $F_3$. The C-series of NLP contains $u=-1$ from $F_2$, $u=s-1$ and then $s=-t$ from $F_2$. The large $t$ tail cancels within the B series. In terms of the general notations in Sec.~\ref{sec:generaltwist}, we have ${\cal H}_B^{(1)}\equiv {\cal H}_1^{1,0}+q_1^{1,0}+{\cal H}_3^{1,0}+q_3^{1,0}$, $\hat q_B^{(1)} \equiv \hat q_1^{(1)}+\hat q_3^{(1)}$, ${\cal H}_C^{(1)} \equiv {\cal H}_2^{1,0}$ and $q_C^{(1)} \equiv q_2^{(1)}$. 

First, one has
\begin{align}\label{eq:qb}
&\hat q^{(1)}_B(t,\lambda)\nonumber \\ 
&=-\frac{1}{4\pi}\int_{c-i\infty}^{c+i\infty} \frac{ds}{2\pi i} \frac{\Gamma(1-2s)\Gamma(s+t)\Gamma(-1-s)\Gamma(4+2s)}{\Gamma(1-s+t)}\,_1\tilde F_1(4+2s,3+s,-\lambda) \nonumber \\ 
&-\frac{\lambda}{2\pi}\int_{c-i\infty}^{c+i\infty} \frac{ds}{2\pi i} \frac{\Gamma(1-2s)\Gamma(s+t)\Gamma(-1-s)\Gamma(4+2s)}{\Gamma(1-s+t)}\,_1\tilde F_1(4+2s,4+s,-\lambda) \ .
\end{align}
It has again the $\frac{1}{t}$ large $t$ tail, but is finite otherwise: 
\begin{align}
\hat q^{(1)}_B(t,\lambda)|_{t\rightarrow \infty}=-\frac{\Gamma(4)}{4\pi}\frac{1}{t}\,_1\tilde F_1(4,3,-\lambda) -\frac{\Gamma(4)}{2\pi}\frac{\lambda}{t}\,_1\tilde F_1(4,4,-\lambda)+{\cal O}\left(\frac{1}{t^2}\right) \ .
\end{align}
To cancel the large $t$ tail, one needs the leading power part of 
\begin{align}
&{\cal F}_B(z^2m^2,t,\lambda)\nonumber \\ 
&=-\frac{1}{4\pi}\int_{c-i\infty}^{c+i\infty} \frac{ds}{2\pi i} \left(\frac{z^2m^2}{4}\right)^{-s} \frac{\Gamma(1-2s)\Gamma(s+t)\Gamma(s-1)\Gamma(4)}{\Gamma(1-s+t)}\,_1\tilde F_1(4,3-s,-\lambda) \nonumber \\ 
&-\frac{\lambda}{2\pi}\int_{c-i\infty}^{c+i\infty} \frac{ds}{2\pi i} \left(\frac{z^2m^2}{4}\right)^{-s} \frac{\Gamma(1-2s)\Gamma(s+t)\Gamma(s-1)\Gamma(4)}{\Gamma(1-s+t)}\,_1\tilde F_1(4,4-s,-\lambda) \ .
\end{align}
Namely
\begin{align}\label{eq:hb}
&{\cal H}^{(1)}_B(t,\alpha(z),\lambda)\nonumber \\ =&-\frac{\Gamma(4)}{4\pi}\bigg(-\frac{1}{t}\,_1\tilde F_1(4,3,-\lambda) +\left(\frac{z^2m^2}{4}\right)^{t}\Gamma(-t-1)\tilde F_1(4,3+t,-\lambda)\bigg) \nonumber \\ 
&-\frac{\Gamma(4)\lambda}{2\pi}\bigg(-\frac{1}{t}\,_1\tilde F_1(4,4,-\lambda) +\left(\frac{z^2m^2}{4}\right)^{t}\Gamma(-t-1)\tilde F_1(4,4+t,-\lambda)\bigg)
\end{align}
Now, the $t=0$ pole cancesl among ${\cal H}^{1}_B(t,\alpha(z),\lambda)$, and the large $t$ tail cancels the one for $\hat q_B^{(1)}(t,\lambda)$. 

Finally, one comes to the series C, generated from the  $F_2$ function. One has
\begin{align}\label{eq:qc}
&q^{(1)}_C(t,\lambda)\nonumber \\ 
&=\frac{1}{\pi}\int_{c-i\infty}^{c+i\infty} \frac{ds}{2\pi i}  \frac{s\Gamma(1-2s)\Gamma(s+t)\Gamma(-s)\Gamma(2+2s)}{\Gamma(1-s+t)}\,_1\tilde F_1(2s+2,3+s,-\lambda) \ .
\end{align}
It is finite and has no large $t$ tail. Similarly, the hard part can be generated through
\begin{align}
&{\cal F}_{C}(z^2m^2,t,\lambda)\nonumber \\ 
&=-\frac{1}{\pi}\int_{c-i\infty}^{c+i\infty} \frac{ds}{2\pi i}  \left(\frac{z^2m^2}{4}\right)^{-s} \frac{s\Gamma(1-2s)\Gamma(s+t)\Gamma(s-1)}{\Gamma(1-s+t)}\,_1\tilde F_1(2,3-s,-\lambda) \ , 
\end{align}
namely,
\begin{align}\label{eq:hc}
{\cal H}^{(1)}_C(t,\alpha(z),\lambda)=\frac{1}{\pi}\left(\frac{z^2m^2}{4}\right)^{t} \frac{\Gamma(-t+1)}{t+1}\,_1\tilde F_1(2,3+t,-\lambda) \ .
\end{align}
It has no singularity at $t=0$.

\subsection{Summary of results up to NLP}
The above finished the enumeration of all the contributions at LP and NLP. To summarize, at LP, there is a hard kernel $-{\cal H}^{(0)}(t,\alpha(z),\lambda)$ in Eq.~(\ref{eq:h0}) and a non-perturbative function $-\hat q_1(t,\lambda)-\hat q_3(t,\lambda)+q_2(t,\lambda)$ in Eq.~(\ref{eq: q1}) and Eq.~(\ref{eq:q2}). The large $\frac{1}{t}$ tail cancels between $\hat q_1(t,\lambda)+\hat q_3(t,\lambda)$ and ${\cal H}^{0}(t,\alpha(z),\lambda)$. At NLP, there are three series of non-perturbative functions and the related hard functions:
\begin{enumerate}
    \item $-q^{(1)}_A(t,\lambda)$ in Eq.~(\ref{eq:qA}) and $-{\cal H}^{(1)}_A(t,\alpha(z),\lambda)$ in Eq.~(\ref{eq:ha}).
    \item  $-\hat q^{(1)}_B(t,\lambda)$ in Eq.~(\ref{eq:qb}) and $-{\cal H}^{(1)}_B(t,\alpha(z),\lambda)$ in Eq.~(\ref{eq:hb}).
    \item $q^{(1)}_C(t,\lambda)$ in Eq.~(\ref{eq:qc}) and ${\cal H}^{(1)}_C(t,\alpha(z),\lambda)$ in Eq.~(\ref{eq:hc}). 
\end{enumerate}
 The $t=1$ renormalon pole cancels between the ${\cal H}^{(0)}(t,\alpha(z),\lambda)$ and $\frac{z^2m^2}{4}q^{(1)}_A(t,\lambda)$. The large $\frac{1}{t}$ tail cancels between $\hat q^{(1)}_B(t,\lambda)$ and ${\cal H}^{(1)}_B(t,\alpha(z),\lambda)$.  More explicitly, one has:
\begin{align}
&{\cal E}(z^2m^2,i\lambda)=\frac{2\pi e^{\lambda}}{N}\int_{0}^{\infty} dt\bigg(-\hat q_1(t,\lambda)-\hat q_3(t,\lambda)+q_2(t,\lambda)-{\cal H}^{(0)}(t,\alpha(z),\lambda) \bigg) \nonumber \\ 
&+\frac{2\pi e^{\lambda}}{N}\frac{z^2m^2}{4}\int_{0}^{\infty} dt \bigg(-q_A^{(1)}(t,\lambda)-{\cal H}^{(1)}_A(t,\alpha(z),\lambda)-\hat q_B^{(1)}(t,\lambda)-{\cal H}^{(1)}_B(t,\alpha(z),\lambda)\bigg) \nonumber \\ 
&+\frac{2\pi e^{\lambda}}{N}\frac{z^2m^2}{4}\int_{0}^{\infty} dt \bigg(q_C^{(1)}(t,\lambda)+{\cal H}^{(1)}_C(t,\alpha(z),\lambda)\bigg)+\frac{1}{N}{\cal O}\left(z^4m^4 \ln \alpha(z) \right) \ .
\end{align}
The above is the first major result of this paper. Notice that although up to now the $\hat q_1(t,\lambda)$, $q_2(t,\lambda)$, and the NLP non-perturbative functions are still in Mellin's representation, explicit forms them can actually be written down. This will be shown in Sec.~\ref{sec:PDF}, see Eq.~(\ref{eq:q1full}) and Eq.~(\ref{eq:q2full}).

Here we would like to comment on the operator structure at NLP. At order $\frac{1}{N}$, in addition to the ``kinematic ones'', one needs at least all the operators below ($n\ge 0$):
\begin{align}
&\bar \psi_i \partial^2 (\partial^+)^n\psi_i \ ,  
\bar \psi_i \gamma^+ \partial^2 (\partial^+)^n\psi_i \bar\psi \psi \ ,\bar \psi_i (\partial^+)^n\psi_i (\bar \psi \psi)^2 \ , \nonumber \\ 
&\bar \psi_i \gamma^+  (\partial^+)^n\psi_i (\bar \psi \psi)^3 \ , \bar \psi_i \gamma^+(\partial^+)^n\psi_i \bar \psi \partial^2\psi  \ .... \ , 
\end{align}
where the additional ``sea-quark operators'' are sensitive to vacuum condensates. In fact, we have seen that even at the leading power, there is a contribution from $g_0^2\langle \bar \psi \psi\rangle$.  As such, the ``internal structure'' of the fermion and the vacuum can not be cleanly separated for our correlator, thus in the strict sense the ``collinear factorization'' in the partonic picture is invalid, but the OPE expansion always works.

\subsection{Scheme/scale dependency of non-perturbative functions and hard kernels}\label{sec:scalescheme}
The fact that the large $\frac{1}{t}$ tails have to cancel between $\hat q_1+\hat q_3$ and ${\cal H}^{(0)}$, $\hat q^{(1)}_B$ and ${\cal H}^{(1)}_B$ implies that these pairs must be combined together to be meaningful. Here we further investigate this issue. 

We introduce for $0<\mu<1$ (which plays the role of ``renormalization scale'')
\begin{align}
{\cal H}_1^{(0)}(t,\alpha(z),\lambda,\mu)=&\frac{1}{4\pi}\left(\frac{z^2m^2}{4}\right)^{t}\bigg(\Gamma(-t)\,_1\tilde F_1(2,1+t,-\lambda)+\frac{1}{t}\,_1\tilde F_1(2,1,-\lambda)\bigg)\nonumber \\ 
&-\frac{1}{4\pi t}\,_1\tilde F_1(2,1,-\lambda)\bigg(\left(\frac{z^2m^2}{4}\right)^{t}-\mu^t\bigg)\ , 
\end{align}
and
\begin{align}
{\cal H}_3^{(0)}(t,\alpha(z),\lambda,\mu)=&\frac{\lambda}{2\pi}\left(\frac{z^2m^2}{4}\right)^{t}\bigg(\Gamma(-t)\,_1\tilde F_1(2,2+t,-\lambda)+\frac{1}{t}\,_1\tilde F_1(2,2,-\lambda)\bigg)\nonumber \\ 
&-\frac{\lambda}{2\pi t}\,_1\tilde F_1(2,2,-\lambda)\bigg(\left(\frac{z^2m^2}{4}\right)^{t}-\mu^t\bigg)\ .
\end{align}
Notice that in terms of the general notations in Sec.~\ref{sec:generaltwist}, they are labeled as  ${\cal H}_1^{0,0}(t,\alpha(z),\lambda,\mu)$ and ${\cal H}_3^{0,0}(t,\alpha(z),\lambda,\mu)$, respectively. Now, one defines 
\begin{align}
&q_1(t,\lambda,\mu)=\hat q_1(t,\lambda)+\frac{1}{4\pi}\frac{1}{t}\,_1\tilde F_1(2,1,-\lambda)(1-\mu^t) \ , \\ 
&q_1(\lambda,\mu)=\int_0^{\infty} dt q_1(t,\lambda,\mu) \ , \label{eq:q1mu}
\end{align}
and
\begin{align}
&q_3(t,\lambda,\mu)=\hat q_3(t,\lambda)+\frac{\lambda}{2\pi}\frac{1}{t}\,_1\tilde F_1(2,2,-\lambda)(1-\mu^t) \ , \\ 
&q_3(\lambda,\mu)=\int_0^{\infty} dt q_3(t,\lambda,\mu) \ . \label{eq:q3mu}
\end{align}
Clearly, the large $t$ tail has all been removed in a way without introducing small $t$ singularities in this manner. Notice that the way to make sense $q_1$ and ${\cal H}^{(0)}$ is not unique. One can replace $\mu^t$ by any analytic function $\chi(t)$ in a neighbourhood of the positive real axis such that $\chi(0)=1$ and $\chi(t)|_{t\rightarrow +\infty}={\cal O}\left(\frac{1}{t}\right)$. Each (family of) such $\chi(t)$ specifies a {\it scheme choice}. 

Due to the renormalon singularities, hard functions after the Borel integrals will always be scheme-dependent. One may ask, is there any scheme independent way to define the non-perturbative functions at leading power? The answer is yes. In fact, in the $z\rightarrow 0$ limit, one has
\begin{align}
{\rm PV}\int_{0}^{\infty} dt {\cal H}_1^{(0)}(t,\alpha(z),\lambda,\mu) \rightarrow \frac{1}{4\pi}\,_1\tilde F_1(2,1,-\lambda) \bigg(\ln \ln \frac{4}{z^2m^2}-\ln \ln \frac{1}{\mu}\bigg)+{\cal O}\left(\frac{1}{\ln \frac{1}{z^2m^2}}\right) \ .
\end{align}
Thus, one can define the ``scale-invariant PDF'' as
\begin{align}\label{eq:scaleq1}
q_1(\lambda)=\int_{0}^{\infty} dt q_1(t,\lambda,\mu)-\frac{1}{4\pi}\,_1\tilde F_1(2,1,-\lambda)\ln \ln \frac{1}{\mu} \ ,
\end{align}
which is $\mu$-independent. Moreover, the leading $z^2\rightarrow 0$ asymptotics reads
\begin{align}
&F_1(z^2m^2,\lambda)|_{z^2\rightarrow 0}\rightarrow \frac{1}{4\pi}\,_1\tilde F_1(2,1,-\lambda) \ln \ln \frac{1}{z^2m^2}+q_1(\lambda)+{\cal O}\left(\alpha(z)\right)  \ , \\
&F_2(z^2m^2,\lambda)|_{z^2\rightarrow 0}\rightarrow q_2(\lambda)+{\cal O}\left(z^2m^2\alpha(z)\right) \ . 
\end{align}
This implies that one can actually {\it define} $q_1(\lambda)$ as 
\begin{align}
q_1(\lambda) \equiv \lim_{z^2\rightarrow 0^+} \bigg( F_1(z^2m^2,\lambda)-\frac{1}{4\pi}\,_1\tilde F_1(2,1,-\lambda) \ln \ln \frac{1}{z^2m^2}\bigg) \ .
\end{align}
Similarly, one can also define 
$q_3(\lambda)$ as
\begin{align}
q_3(\lambda) \equiv \lim_{z^2\rightarrow 0^+} \bigg( F_3(z^2m^2,\lambda)-\frac{\lambda}{2\pi}\,_1\tilde F_1(2,2,-\lambda) \ln \ln \frac{1}{z^2m^2}\bigg) \ .
\end{align}
In this way, the scale and scheme independence of  $q_1(\lambda)$ and $q_3(\lambda)$ is manifest. Explicit form of the scale-invariant ``PDF'' $q_1$ will be given in Eq.~(\ref{eq:q1full}) and $q_2$ in Eq.~(\ref{eq:q2full}). Similar construction applies to $\hat q^{(1)}_B$ and ${\cal H}^{(1)}_B$ as well.

\subsection{Simplifying non-perturbative functions and the ``small-$x$'' limit}\label{sec:PDF}
In fact, the non-perturbative functions can be further simplified in the momentum-fraction space, up to subtleties caused by the subtractions at $\alpha=0, 1$. In this subsection we pursue this direction for all the non-perturbative functions and the leading twist.

We first consider the $q_1(t,\lambda)$. Naively, one can consider the representations
\begin{align}
&e^{\lambda}\,_1\tilde F_1(2s+2,1+s,-\lambda)=\frac{1}{\Gamma(2s+2)\Gamma(-s-1)}\int_{0}^1 e^{\alpha \lambda}(1-\alpha)^{2s+1}\alpha^{-2-s} d\alpha \ , \\ 
&e^{\lambda}\,_1\tilde F_1(2s+2,2+s,-\lambda)=\frac{1}{\Gamma(2s+2)\Gamma(-s)}\int_{0}^1 e^{\alpha \lambda}(1-\alpha)^{2s+1}\alpha^{-1-s} d\alpha \ ,
\end{align}
and 
\begin{align}
e^{\lambda}\,_1\tilde F_1(2s,1+s,-\lambda)=\frac{1}{\Gamma(2s)\Gamma(-s+1)}\int_{0}^1 e^{\alpha \lambda }(1-\alpha)^{2s-1}\alpha^{-s}d\alpha \ .
\end{align}
Neglecting the issue related to subtraction terms at $\alpha=0$ and $\alpha=1$, one can proceed to
\begin{align}
& (e^{\lambda}q_1)(t,\alpha)=-\frac{\alpha^{-2}(1-\alpha)}{4\pi}\int_{c-i\infty}^{c+i\infty} \frac{ds}{2\pi i}\frac{(s+1)\Gamma(1-2s)\Gamma(s+t)}{\Gamma(1-s+t)}(1-\alpha)^{2s}\alpha^{-s} \ , \\ 
& \left(\frac{1}{\lambda}e^{\lambda}q_3\right)(t,\alpha)=\frac{\alpha^{-1}(1-\alpha)}{2\pi}\int_{c-i\infty}^{c+i\infty} \frac{ds}{2\pi i}\frac{\Gamma(1-2s)\Gamma(s+t)}{\Gamma(1-s+t)}(1-\alpha)^{2s}\alpha^{-s} \ , \\ 
&(e^{\lambda}q_2)(t,\alpha)=-\frac{(1-\alpha)^{-1}}{\pi}\int_{c-i\infty}^{c+i\infty} \frac{ds}{2\pi i}\frac{s\Gamma(1-2s)\Gamma(s+t)}{\Gamma(1-s+t)}(1-\alpha)^{2s}\alpha^{-s} \ .
\end{align}
Now, integrating by shifting the contour to the left one obtains after summing over residues
\begin{align}
&(e^{\lambda}q_1)(t,\alpha)=-\frac{(1-\alpha)^2 \alpha^{t-2} \left(t (\alpha^2-1)+1+4\alpha+\alpha^2\right)}{4\pi(\alpha+1)^3} \ , \\
&\left(\frac{1}{\lambda}e^{\lambda}q_3\right)(t,\alpha)=\frac{(1-\alpha)^2\alpha^{-1+t}}{2\pi(1+\alpha)} \ , \\ 
&(e^{\lambda}q_2)(t,\alpha)=-\frac{\alpha^t(t(\alpha^2-1)+2\alpha)}{\pi(\alpha+1)^3} \ .
\end{align}
After integrating over $t$, one obtains
\begin{align}
&(e^{\lambda}q_1)(\alpha)=\frac{(1-\alpha)^2 \left(1-\alpha^2+(1+4\alpha+\alpha^2)\ln \alpha\right)}{4\pi \alpha^2 (1+\alpha)^3 \ln ^2\alpha} \ , \\ 
&\left(\frac{1}{\lambda}e^{\lambda}q_3\right)(\alpha)=-\frac{(1-\alpha)^2}{2\pi \alpha(1+\alpha)\ln \alpha} \ , \\ 
&(e^{\lambda}q_2)(\alpha)=\frac{1-\alpha^2+2\alpha \ln \alpha}{\pi (1+\alpha)^3\ln^2 \alpha} \ ,
\end{align}
up to subtractions at $\alpha=0$ and $\alpha=1$. Here we must determine them by matching to the coordinate space asymptotics.

We first consider the most singular one, the $q_1$. In the $\alpha \rightarrow 1$ limit, one has $q_1(\alpha) \sim -\frac{1-\alpha}{8\pi}$, consistent with the $-\frac{1}{8\pi \lambda^2}$ asymptotics. Thus, no subtraction term at $\alpha=1$ is required. Thus, one needs to determine the subtraction term at $\alpha=0$. For this purpose, notice that one always has
\begin{align}
&(e^{\lambda}q_1)(\alpha)-\bigg(\frac{\left(1-\alpha\right)^2}{4\pi \alpha^2\ln ^2\alpha}\bigg)_{+}\frac{1-\alpha^2}{(1+\alpha)^3}-\bigg(\frac{(1-\alpha)^2}{4\pi \alpha^2 \ln \alpha}\bigg)_{+}\frac{1+4\alpha+\alpha^2}{(1+\alpha)^3}
\nonumber \\ 
&=a_1\delta(\alpha)+a_2\delta'(\alpha) \ ,
\end{align}
where the ``plus'' distributions is defined as subtracting $\psi(0)+\alpha \psi'(0)$ for any test function $\psi(\alpha)$.  Clearly, acting on test functions for which $\psi^{k}(0)=0$ for all $k\ge 0$, the subtraction has no effect. One now needs to fix the $a_1$ and $a_2$. For this purpose, we consider the $\lambda \rightarrow \infty$ asymptotics in coordinate space for the function
\begin{align}
e^{-\lambda}q_1(-\lambda,\mu)=e^{\lambda}\int_{0}^{\infty} dt \bigg(\hat q_1(t,-\lambda)+\frac{1}{4\pi}\frac{1}{t}\,_1\tilde F_1(2,1,\lambda)(1-\mu^t)\bigg) \ ,
\end{align}
and compare with that for the function
\begin{align}
&Q(-\lambda)\nonumber \\ 
&=\int_{0}^{1} d\alpha \bigg(\frac{\left(1-\alpha\right)^2}{4\pi \alpha^2\ln ^2\alpha}\bigg)_{+}\frac{1-\alpha^2}{(1+\alpha)^3}e^{-\lambda \alpha}+\int_{0}^{1} d\alpha \bigg(\frac{\left(1-\alpha\right)^2}{4\pi \alpha^2\ln \alpha}\bigg)_{+}\frac{1+4\alpha+\alpha^2}{(1+\alpha)^3}e^{-\lambda \alpha} \ . 
\end{align}
This will determine the unknown constants $a_1$ and $a_2$. Now, using the chain rule of the derivatives, it is easy to show that the large $\lambda$ asymptotics, up to order ${\cal O}(\lambda^0)$, is controlled by 
\begin{align}
{\cal G}(\lambda)=\int_{0}^{1} d\alpha \frac{(1-\alpha)^2(1+\ln \alpha)}{4\pi \alpha^2 \ln^2 \alpha}\left(e^{-\lambda \alpha}-1+\lambda \alpha\right) \ ,
\end{align}
and
\begin{align}
{\cal R}(\lambda)=\int_{0}^{1} d\alpha \frac{(1-\alpha)^2(\ln \alpha-3)}{4\pi \alpha \ln^2 \alpha}\left(e^{-\lambda \alpha}-1\right) \ .
\end{align}
We now consider the large $\lambda$ asymptotics for the above. For ${\cal R}(\lambda)$, one has 
\begin{align}
{\cal R}(\lambda)\rightarrow  \frac{1}{4\pi}\left(\gamma_E+5\ln 2+\ln \ln \lambda\right)+{\cal O}\left(\frac{1}{\ln \lambda}\right) \ .
\end{align}
Similarly, one has for ${\cal G}(\lambda)$
\begin{align}
{\cal G}(\lambda) \rightarrow &\frac{\lambda}{4\pi}\left(-\gamma_E+3\ln 2-\ln \ln \lambda+{\cal O}\left(\frac{1}{\ln \lambda}\right)\right)\nonumber \\ 
&-\frac{1}{2\pi}\left(\gamma_E+\ln \ln \lambda+{\cal O}\left(\frac{1}{\ln \lambda}\right)\right) \ .
\end{align}
The above determines the leading coefficients for the large $-\lambda$ asymptotics from momentum space representation for $Q(\lambda)$
\begin{align}
Q(-\lambda) \rightarrow \frac{\lambda+1}{4\pi}\bigg(-\gamma_E+3\ln 2-\ln \ln \lambda\bigg) +\frac{\ln 2}{2\pi} \ .
\end{align}
Similarly, one can directly consider the expansion of $\hat q_1(t,-\lambda)$ in the large $\lambda$ limit.   From the expansion of the hypergeometric functions, one has
\begin{align}
e^{-\lambda}\hat q_1(t,-\lambda) \sim \frac{1}{4\pi}\int_{c-i\infty}^{c+i\infty}\frac{ds}{2\pi i}\frac{\Gamma(1-2s)\Gamma(s+t)\Gamma(-s)}{\Gamma(1-s+t)}\lambda^{1+s} \nonumber \\ 
+\frac{1}{4\pi}\int_{c-i\infty}^{c+i\infty}\frac{ds}{2\pi i}\frac{(s+1)(2s+1)\Gamma(1-2s)\Gamma(s+t)\Gamma(-s)}{\Gamma(1-s+t)}\lambda^{s} \ .
\end{align}
By shifting the contour to the left, one has for the scale-dependent $q_1$ Eq.~(\ref{eq:q1mu})
\begin{align}
e^{-\lambda}q_1(-\lambda,\mu) \rightarrow \frac{\lambda+1}{4\pi}\left(\ln \ln \frac{1}{\mu}-\ln \ln \lambda\right) \ .
\end{align}
From the above and Eq.~(\ref{eq:scaleq1}), one obtains the difference 
\begin{align}
\int_{0}^1 d\alpha e^{-\lambda \alpha}\left(a_1\delta(\alpha)+a_2\delta'(\alpha)\right)=\frac{\lambda+1}{4\pi}\bigg(\gamma_E-3\ln 2\bigg) -\frac{\ln 2}{2\pi} \ .
\end{align}
This completely determines the scale-invariant non-perturbative function $q_1$:
\begin{align}\label{eq:q1full}
&e^{-\lambda}q_1(-\lambda)\nonumber \\ 
&=\int_{0}^{1} d\alpha \bigg(\frac{\left(1-\alpha\right)^2}{4\pi \alpha^2\ln ^2\alpha}\bigg)_{+}\frac{1-\alpha^2}{(1+\alpha)^3}e^{-\lambda \alpha}+\int_{0}^{1} d\alpha \bigg(\frac{\left(1-\alpha\right)^2}{4\pi \alpha^2\ln \alpha}\bigg)_{+}\frac{1+4\alpha+\alpha^2}{(1+\alpha)^3}e^{-\lambda \alpha} \nonumber \\ 
& + \frac{\lambda+1}{4\pi}\bigg(\gamma_E-3\ln 2\bigg) -\frac{\ln 2}{2\pi} \ .
\end{align}
Similarly, for $q_2$ one can see that there is no subtraction term at $\alpha=0$, but one needs the subtraction term $-\frac{e^{-\lambda}}{4\pi}$ at $\alpha=1$, which should be related to over-all normalization of the 
correlator. Thus one has
\begin{align}\label{eq:q2full}
e^{-\lambda}q_2(-\lambda)=\int_{0}^1 d\alpha \frac{1-\alpha^2+2\alpha \ln \alpha}{\pi(1+\alpha)^3\ln^2 \alpha} e^{-\lambda \alpha}-\frac{e^{-\lambda}}{4\pi} \ .
\end{align}
For $q_3$, there is a subtraction at $\alpha=0$, but no subtraction at $\alpha=1$, and one has
\begin{align}
e^{-\lambda}q_3(-\lambda)=\lambda\int_{0}^{1} d\alpha \bigg(\frac{(1-\alpha)^2}{2\pi \alpha \ln \alpha}\bigg)_+\frac{e^{-\alpha \lambda}}{1+\alpha}-\frac{\gamma_E-\ln 2}{2\pi}\lambda \ .
\end{align}
As far as we know, this is the first time in the literature that a non-perturbative scaling function appearing in the Bjorken limit has been fully determined to order $\frac{1}{N}$ in 2D models. The non-perturbative functions $q_B^{(1)}$ and $q_C^{(1)}$ at NLP can also be worked out explicitly. We should mention that for non-perturbative functions $q^{(l)}(\lambda,\mu)$ at higher powers with $l\ge 2$ introduced latter in Sec.~\ref{sec:generaltwist}, there are additional poles from ratios of gamma functions such as $\frac{\Gamma(-l-s)}{\Gamma(-s-1)}$ for $q_1^{(l)}$ in Eq.~(\ref{eq:q1lmu}) and the treatment for them are more tricky.

\section{Twist expansion and renormalon cancellation to all powers}\label{sec:generaltwist}
After discussing in detail the twist expansion up to NLP, in this section we investigate the twist expansion and renormalon cancellation to all powers. For demonstration purposes, we only include details of the derivations for the $F_1$, and present the final results for $F_3$ and $F_2$.

\subsection{Twist expansion of $F_1(z^2m^2,\lambda)$}
We first introduce the set of non-perturbative functions at powers $\left(\frac{z^2m^2}{4}\right)^l$. They are represented as
\begin{align}\label{eq:q1l}
\hat q_1^{(l)}(t,\lambda)=\frac{(-1)^l}{4\pi l!}\int_{c-i\infty}^{c+i\infty}\frac{ds}{2\pi i}&\frac{\Gamma(-l-s)\Gamma(2l+2+2s)\Gamma(1-2s)\Gamma(s+t)}{\Gamma(1-s+t)}\nonumber \\ & \times\,_1\tilde F_1(2+2l+2s,1+2l+s,-\lambda) \ .
\end{align}
They have no singularities for $t\ge 0$, but contain large $t$ tails
\begin{align}
\hat q^{(l)}_1(t,\lambda)|_{t\rightarrow \infty} \rightarrow -\frac{\Gamma(2l+2)}{4\pi t (l!)^2}\,_1\tilde F_1(2+2l,1+2l,-\lambda) \ .
\end{align}
One then introduces the ``scale-dependent'' non-perturbative functions
\begin{align}
&q_1^{(l)}(t,\lambda,\mu)=\hat q_1^{(l)}(t,\lambda)+\frac{\Gamma(2l+2)}{4\pi (l!)^2}\,_1\tilde F_1(2+2l,1+2l,-\lambda)\frac{1-\mu^t}{t} \ , \\
&q_1^{(l)}(\lambda,\mu)=\int_{0}^{\infty} dt q_1^{(l)}(t,\lambda,\mu) \ , \label{eq:q1lmu}
\end{align}
which are well-defined. 

Now, one introduces the ``mother functions'' for perturbative kernels at {\it level} $l$ 
\begin{align}\label{eq:motherlF1}
{\cal F}_1^{(l)}(z^2m^2,t,\lambda)=\frac{(-1)^l\Gamma(2l+2)}{4\pi l!}&\int_{c-i\infty}^{c+i\infty}\frac{ds}{2\pi i}\frac{\Gamma(-l+s)\Gamma(1-2s)\Gamma(s+t)}{\Gamma(1-s+t)}\nonumber \\ & \times \,_1\tilde F_1(2+2l
,1+2l-s,-\lambda)\left(\frac{z^2m^2}{4}\right)^{-s} \ ,
\end{align}
as well as the scale-dependent ones, which subtracts out the large $t$ tail:
\begin{align}
{\cal F}_1^{(l)}(z^2m^2,t,\lambda,\mu)={\cal F}_1^{(l)}(z^2m^2,t,\lambda)-\frac{\Gamma(2l+2)}{4\pi (l!)^2}\,_1\tilde F_1(2+2l,1+2l,-\lambda)\frac{1-\mu^t}{t} \ , \\ 
{\cal F}_1^{(l)}(z^2m^2,\lambda,\mu)=\int_{0}^{\infty} dt {\cal F}_1^{(l)}(z^2m^2,t,\lambda,\mu)  \ .\label{eq:motherF1mu}
\end{align}
Notice that these $\mu$-dependent mother functions contain no singularities at $t\ge 0$ and integrate over $t$ to finite functions. Now, by picking up poles from $s=-p$ and $s=-t-p$ with $p\ge 0$, one obtains a series of non-perturbative functions as well as hard kernels. Due to the fact that there are initially no singularities in the positive real axis $t\ge 0$ at all, all the generated singularities must cancel between the hard functions and non-perturbative functions:
\begin{align}
{\cal F}_1^{(l)}(z^2m^2,t,\lambda)=\sum_{p=0}^{\infty}\left(\frac{z^2m^2}{4}\right)^{p}\bigg({\cal H}_1^{l,p}(t,\alpha(z),\lambda)+q_{1}^{l,p}(t,\lambda)\bigg) \ ,
\end{align}
where one has the hard kernels
\begin{align}\label{eq:H1lp}
{\cal H}_1^{l,p}(t,\alpha(z),\lambda)= &\frac{(-1)^{l+p} \Gamma (2 l+2) \Gamma (2t+2p+1) \Gamma (-l-p-t)}{4 \pi  l! p!  \Gamma (2 t+p+1)} \nonumber \\ \times &\,_1\tilde F_1(2+2l,1+2l+p+t,-\lambda)\left(\frac{z^2m^2}{4}\right)^t \ , 
\end{align}
and the non-perturbative functions
\begin{align}\label{eq:q1lp}
q_1^{l,p}(t,\lambda)=\frac{(-1)^{p} \Gamma (2 l+2) \Gamma (2 p+1) }{4 \pi  l! (l+p)! }\frac{\Gamma(t-p)}{\Gamma(t+p+1)}\,_1\tilde F_1(2+2l,1+2l+p,-\lambda) \ .
\end{align}
Notice that the ``non-perturbative functions'' with $p=0$ are just the 
$\frac{1}{t}$ tails for the ${\cal F}^{(l)}(z^2m^2,t,\lambda)$ functions, which also cancel the $t=0$ singularities for ${\cal H}^{l,0}_1$, therefore can be assigned to the hard kernels ${\cal H}_1^{l,0}$ as well. In particular, in the previous discussions the $q_1^{0,0}$ and $q_1^{1,0}$ are assigned to the hard kennels ${\cal H}^{(0)}$ and ${\cal H}^{(1)}_B$.

Notice that for any $p$,  there are only $t=0,...p$  poles for $q_1^{l,p}(t,\lambda)$. Thus one must show that the poles of ${\cal H}_1^{l,p}(t,\alpha(z),\lambda)$ at $t=n$ ($n \ge 0$)  cancel the ones for $\left(\frac{z^2m^2}{4}\right)^nq^{l,p+n}_1(t,\lambda)$ at $t=n$. Indeed, one has
\begin{align}
{\rm Res} {\cal H}_{1}^{l,p}(t=n,\alpha(z),\lambda)=&\frac{(-1)^{l+p}\Gamma(2l+2)\Gamma(2n+2p+1)}{4\pi l!p!\Gamma(2n+p+1)}\frac{(-1)^{l+n+p+1}}{(l+n+p)!}\left(\frac{z^2m^2}{4}\right)^n \nonumber \\ 
&\times \,_1\tilde F_1(2+2l,1+2l+p+n,-\lambda) \ ,
\end{align}
while one has
 \begin{align}
\left(\frac{z^2m^2}{4}\right)^n {\rm Res} q_{1}^{l,p+n}(t=n,\lambda)=&\frac{(-1)^{p+n} \Gamma (2 l+2) \Gamma (2 p+2n+1) }{4 \pi  l! (l+p+n)! }\frac{(-1)^p}{p!\Gamma(2n+p+1)}\nonumber \\ 
&\times \,_1\tilde F_1(2+2l,1+2l+p+n,-\lambda)\left(\frac{z^2m^2}{4}\right)^n \ ,
 \end{align}
which indeed cancels the pole of ${\cal H}_1^{l,p}(t,\alpha(z),\lambda)$ at $t=n$. As a result, we have shown that all the renormalon poles obtained by expanding ${\cal F}_{1}^{(l)}(z^2m^2,t,\lambda,\mu)$ completely cancel out. Notice that for $n=0$,  the $t=0$ poles cancel between the hard and non-perturbative contributions at the same power, suggesting that one can also introduce the corresponding $\mu$ dependent versions for these functions. In particular, for $p=0$ one can introduce 
\begin{align}
&{\cal H}_1^{l,0}(t,\alpha(z),\lambda,\mu) \equiv \nonumber \\ 
& {\cal H}_1^{l,0}(t,\alpha(z),\lambda)+q_1^{l,0}(t,\lambda)-\frac{\Gamma(2l+2)}{4\pi (l!)^2}\,_1\tilde F_1(2+2l,1+2l,-\lambda)\frac{1-\mu^t}{t} \ , 
\end{align}
and for $p\ge 1$ such construction applies as well, leading to scale-dependent functions ${\cal H}_1^{l,p}(t,\alpha(z),\mu)$ and $q_1^{l,p}(t,\lambda,\mu)$:
\begin{align}\label{eq:H1lpmu}
&{\cal H}_1^{l,p}(t,\alpha(z),\lambda,\mu)\nonumber \\ 
&\equiv {\cal H}_1^{l,p}(t,\alpha(z),\lambda)+\frac{\mu^t}{4\pi t}\frac{\Gamma(2l+2)\Gamma(2p+1)}{l! (p!)^2(l+p)!}\,_1\tilde F_1(2+2l,1+2l+p,-\lambda) \ , 
\end{align}
and 
\begin{align}\label{eq:q1lpmu}
q_1^{l,p}(t,\lambda,\mu)\equiv q_1^{l,p}(t,\lambda) -\frac{\mu^t}{4\pi t}\frac{\Gamma(2l+2)\Gamma(2p+1)}{l! (p!)^2(l+p)!}\,_1\tilde F_1(2+2l,1+2l+p,-\lambda)  \ .
\end{align}
Notice this construction does not affect the singularities at any finite value of $t>0$.  

To summarize, the full twist expansion of the function $F_1$ reads
\begin{align}\label{eq:F1fullexpansion}
F_1(z^2m^2,\lambda)=&\sum_{l=0}^{\infty} \left(\frac{z^2m^2}{4}\right)^l q_1^{(l)}(\lambda,\mu)\nonumber \\ 
+&\sum_{l=0}^{\infty} \left(\frac{z^2m^2}{4}\right)^l\int_{0}^{\infty} dt \sum_{p=0}^{\infty}\left(\frac{z^2m^2}{4}\right)^{p}\bigg({\cal H}_1^{l,p}(t,\alpha(z),\lambda,\mu)+q_{1}^{l,p}(t,\lambda,\mu)\bigg) \ ,
\end{align}
with $q_1^{l,0}(t,\lambda,\mu) \equiv 0$ in our convention. The renormalon singularity at $t=n$,  cancels for each $l$ between ${\cal H}_1^{l,p}(t,\alpha(z),\lambda,\mu)$ and $q_1^{l,p+n}(t,\lambda,\mu)$ for any $p \ge  0$. The $\mu$ dependencies cancel between $q_1^{(l)}(t,\lambda,\mu)$ and ${\cal H}_1^{l,0}(t,\alpha(z),\lambda,\mu)$, as well as between ${\cal H}_1^{l,p}(t,\alpha(z),\lambda,\mu)$ and $q_1^{l,p}(\lambda,\mu)$ with $p \ge 1$. Furthermore, in Eq.~(\ref{eq:F1fullexpansion}), the integration path can be chosen to be $t=(1 \pm i\epsilon)r$ with $r\ge 0$ or the averages of the two signs (principal value) in order to fully specify each power. The contributions at individual powers are affected by the $\pm i\epsilon$ choices, but the total result remains the same due to the rernormalon cancellation pattern discussed above. 

\subsection{Twist expansion of $F_2(z^2m^2,\lambda)$}
After discussing in detail the twist expansion for $F_1$, we present the results for $F_2$. The full twist expansion of $F_2$ reads
\begin{align}\label{eq:F2fullexpansion}
F_2(z^2m^2,\lambda)=&\sum_{l=0}^{\infty} \left(\frac{z^2m^2}{4}\right)^l q_2^{(l)}(\lambda)\nonumber \\ 
+&\sum_{l=1}^{\infty} \left(\frac{z^2m^2}{4}\right)^l\int_{0}^{\infty} dt \sum_{p=0}^{\infty}\left(\frac{z^2m^2}{4}\right)^{p}\bigg({\cal H}_2^{l,p}(t,\alpha(z),\lambda,\mu)+q_{2}^{l,p}(t,\lambda,\mu)\bigg) \ ,
\end{align}
where one has
\begin{align}\label{eq:q2l}
q_2^{(l)}(t,\lambda)=\frac{(-1)^{l-1}}{l!\pi}\int_{c-i\infty}^{c+i\infty}&\frac{ds}{2\pi i}\frac{s\Gamma(2s+2l)\Gamma(-l+1-s)\Gamma(1-2s)\Gamma(s+t)}{\Gamma(1-s+t)} \nonumber \\ 
&\times \,_1\tilde F_1(2s+2l,1+2l+s,-\lambda) \ , 
\end{align}
which has no large $t$ tail, in terms of which one has
\begin{align}
q_2^{(l)}(\lambda)=\int_{0}^{\infty}dt q_2^{(l)}(t,\lambda)  \ .
\end{align}
For the $l,p$ series, one has 
\begin{align}\label{eq:q2lp}
q_2^{l,p}(t,\lambda)=\frac{p(-1)^{p-1}\Gamma(2l)\Gamma(2p+1)}{\pi(l-1)!(l+p)!}\frac{\Gamma(t-p)}{\Gamma(1+p+t)}\,_1\tilde F_1(2l,1+2l+p,-\lambda) \ ,
\end{align}
which vanishes for $p=0$. The hard kernels are
\begin{align}\label{eq:H2lp}
{\cal H}_2^{l,p}(t,\alpha(z),\lambda)=&\frac{(-1)^{l+p-1}(t+p)\Gamma(2l)\Gamma(1+2t+2p)\Gamma(-t-p-l)}{\pi (l-1)!p!\Gamma(1+2t+p)}\nonumber \\ &\times \,_1 \tilde F_1(2l,1+2l+p+t,-\lambda)\left(\frac{z^2m^2}{4}\right)^t \ .
\end{align}
 One can check in a way similar to $F_1$ that the $t=n$ renormalon poles cancel for each $l$, between ${\cal H}_{2}^{l,p}(t,\alpha(z),\lambda)$ and $\left(\frac{z^2m^2}{
4}\right)^nq^{l,p+n}_2(t,\lambda)$. Scale-dependent versions can also be introduced using previous constructions as 
\begin{align}\label{eq:H2lpmu}
&{\cal H}_2^{l,p}(t,\alpha(z),\lambda,\mu)\nonumber \\ 
&\equiv {\cal H}_2^{l,p}(t,\alpha(z),\lambda)-\frac{\mu^t}{\pi t}\frac{p\Gamma(2l)\Gamma(2p+1)}{(l-1)!(p!)^2(p+l)!} \,_1\tilde F_1(2l,1+2l+p,-\lambda) \ , 
\end{align}
and 
\begin{align}\label{eq:q2lpmu}
q_2^{l,p}(t,\lambda,\mu)\equiv q_2^{l,p}(t,\lambda)+\frac{\mu^t}{\pi t}\frac{p\Gamma(2l)\Gamma(2p+1)}{(l-1)!(p!)^2(p+l)!} \,_1\tilde F_1(2l,1+2l+p,-\lambda) \ .
\end{align}
 Notice that for $p=0$ the subtraction terms all vanish and ${\cal H}_2^{l,0}$, $q_2^{l,0}$ and $q_2^{(l)}$ are automatically scale-invariant. 

\subsection{Twist expansion of $F_3(z^2m^2,\lambda)$}
The expansion for $F_3$ is essentially the same as that for $F_1$. In particular, one has the non-perturbative functions
\begin{align}\label{eq:q3l}
\hat q_3^{(l)}(t,\lambda)=\frac{(-1)^l\lambda}{2\pi l!}\int_{c-i\infty}^{c+i\infty}\frac{ds}{2\pi i}&\frac{\Gamma(-l-s)\Gamma(2l+2+2s)\Gamma(1-2s)\Gamma(s+t)}{\Gamma(1-s+t)}\nonumber \\ & \times \,_1\tilde F_1(2+2l+2s,2+2l+s,-\lambda) \ ,
\end{align}
and their scale-dependent versions
\begin{align}
&q_3^{(l)}(t,\lambda,\mu)=\hat q_3^{(l)}(t,\lambda)+\frac{\Gamma(2l+2)\lambda}{2\pi (l!)^2}\,_1\tilde F_1(2+2l,2+2l,-\lambda)\frac{1-\mu^t}{t} \ , \\
&q_3^{(l)}(\lambda,\mu)=\int_{0}^{\infty} dt q_3^{(l)}(t,\lambda,\mu) \ . \label{eq:q3lmu}
\end{align}
For the $l,p$ series, one has the ``bare'' ones
\begin{align}\label{eq:H3lp}
{\cal H}_3^{l,p}(t,\alpha(z),\lambda)= &\frac{(-1)^{l+p}\lambda \Gamma (2 l+2) \Gamma (2t+2p+1) \Gamma (-l-p-t)}{2 \pi  l! p!  \Gamma (2 t+p+1)} \nonumber \\ \times & \,_1\tilde F_1(2+2l,2+2l+p+t,-\lambda)\left(\frac{z^2m^2}{4}\right)^t \ , 
\end{align}
and 
\begin{align}\label{eq:q3lp}
q_3^{l,p}(t,\lambda)=\frac{(-1)^{p} \lambda \Gamma (2 l+2) \Gamma (2 p+1) }{2\pi  l! (l+p)! }\frac{\Gamma(t-p)}{\Gamma(t+p+1)}\,_1\tilde F_1(2+2l,2+2l+p,-\lambda) \ .
\end{align}
For the scale-dependent ones, for $p=0$ one has
\begin{align}
&{\cal H}_3^{l,0}(t,\alpha(z),\lambda,\mu)\equiv \nonumber \\ 
&{\cal H}_3^{l,0}(t,\alpha(z),\lambda)+q_3^{l,0}(t,\lambda)-\frac{\Gamma(2l+2)\lambda}{2\pi (l!)^2}\,_1\tilde F_1(2+2l,2+2l,-\lambda)\frac{1-\mu^t}{t} \ , 
\end{align}
while for $p\ge 1$ one has
\begin{align}\label{eq:H3lpmu}
&{\cal H}_3^{l,p}(t,\alpha(z),\lambda,\mu)\nonumber \\ 
&\equiv {\cal H}_3^{l,p}(t,\alpha(z),\lambda)+\frac{\mu^t \lambda}{2\pi t}\frac{\Gamma(2l+2)\Gamma(2p+1)}{l! (p!)^2(l+p)!}\,_1\tilde F_1(2+2l,2+2l+p,-\lambda) \ , 
\end{align}
and 
\begin{align}\label{eq:q3lpmu}
q_3^{l,p}(t,\lambda,\mu)\equiv q_3^{l,p}(t,\lambda) -\frac{\mu^t\lambda}{2\pi t}\frac{\Gamma(2l+2)\Gamma(2p+1)}{l! (p!)^2(l+p)!}\,_1\tilde F_1(2+2l,2+2l+p,-\lambda)  \ .
\end{align}
In terms of the above, one has the full expansion for $F_3$:
\begin{align}\label{eq:F3fullexpansion}
F_3(z^2m^2,\lambda)=&\sum_{l=0}^{\infty} \left(\frac{z^2m^2}{4}\right)^l q_3^{(l)}(\lambda)\nonumber \\ 
+&\sum_{l=0}^{\infty} \left(\frac{z^2m^2}{4}\right)^l\int_{0}^{\infty} dt \sum_{p=0}^{\infty}\left(\frac{z^2m^2}{4}\right)^{p}\bigg({\cal H}_3^{l,p}(t,\alpha(z),\lambda,\mu)+q_{3}^{l,p}(t,\lambda,\mu)\bigg) \ .
\end{align}
As the case of $F_1$, the $\mu$ dependency cancels between $q_3^{(l)}(t,\lambda,\mu)$ and ${\cal H}_3^{l,0}(t,\alpha(z),\lambda,\mu)$ for each $l \ge 0$, as well as between ${\cal H}_3^{l,p}(t,\alpha(z),\lambda,\mu)$ and $q_3^{l,p}(t,\lambda,\mu)$ for each $l \ge 0$ and $p\ge 1$. The $t=n$, $n\ge 1$ renormalon pole cancels for each $l \ge 0$, each $p\ge 0$, between ${\cal H}_{3}^{l,p}(t,\alpha(z),\lambda,\mu)$ and $q_3^{l,p+n}(t,\lambda,\mu)$. This finishes the discussion on $F_3$. 

Given $F_1(z^2m^2,\lambda)$, $F_2(z^2m^2,\lambda)$ and $F_3(z^2m^2,\lambda)$, one has for the structure function ${\cal E}$ at order $\frac{1}{N}$
\begin{align}
{\cal E}(z^2m^2,i\lambda)\equiv \frac{2\pi e^{\lambda}}{N}\left(-F_1(z^2m^2,\lambda)+F_2(z^2m^2,\lambda)-F_3(z^2m^2,\lambda)\right) \ .
\end{align}
This finally ends our discussions on the general form of power expansion to all powers and the general pattern of renormalon cancellation, and it serves as the major result of this work. 

\section{Absolute convergence of the twist expansion}
In this section,  we show that the (coordinate-space) twist expansion for $F_1$ is convergent. Convergence for the expansions of $F_2$ and $F_3$ can be shown in similar ways. Notice that the convergence of OPE is a crucial feature of CFTs and holds in the distributional sense in CFTs as well~\cite{Mack:1976pa}. In theories with marginal UV limits, due to the weak decoupling between UV and IR, OPE coefficients attached to operators at various powers need to be defined through Borel integrals of the form 
\begin{align}
I(\xi)={\rm PV}\int_{0}^{\infty} dt \left(-\xi^2+i\xi^0\epsilon \right)^t m^{2t} e^{\frac{\beta_1}{\beta_0}t\ln t} B(t) \ .
\end{align}
In this case, even if the Borel integral converges for any $-\xi^2>0$ and defines smooth ``Schwinger functions'', it may still grow fast at large separations and fail to define tempered distributions. In particular, it may not allow Fourier transforms at all. The convergence discussed in this section is always point-wise, namely, absolute convergence for any fixed $z^2>0$ and $\lambda \ge 0$.  

\subsection{Expansion in $q_1^{(l)}$ and ${\cal F}_1^{(l)}$}
We first show a slightly simpler version, namely, the expansion of $F_1$ in terms of $q_1^{(l)}(\lambda,\mu)$ defined in Eq.~(\ref{eq:q1lmu}) and the mother functions ${\cal F}^{(l)}_1(z^2m^2,\lambda,\mu)$ in Eq.~(\ref{eq:motherF1mu}), is in fact absolutely convergent for any $z^2>0$, $\lambda \ge 0$ and $0<\mu<1$. The crucial observations are actually the following inequalities (under conditions $l\in Z_{\ge 0}$, $0<c<\frac{1}{2}$, $y\in R$, $\lambda \ge 0$) : 
\begin{align}
&\bigg|\,_1F_1(2+2l+2(c+iy),1+2l+(c+iy),-\lambda)\bigg| \le 1 \ , \\ 
&\bigg|\,_1F_1(2+2l,1+2l-(c+iy),-\lambda)\bigg| \le 1\ , \\ 
&\bigg|\frac{\Gamma(-l+c+iy)}{\Gamma(1+2l-c-iy)}\bigg| \le \bigg|\frac{\Gamma(-l+c)}{\Gamma(1+2l-c)}\bigg| \ , \\ 
&\bigg|\frac{\Gamma(2c+2iy+2l+2)}{\Gamma(1+2l+c+iy)}\bigg|\le \bigg|\frac{\Gamma(2l+2+2c)}{\Gamma(1+2l+c)} \bigg|\ .
\end{align}
To proceed, we shift the $t$ integral to $(1,\infty)$ using the change of variable $t \rightarrow t+1$, and introduce the remainder function
\begin{align} \label{eq:remainder}
&r(s,t)=\frac{\Gamma(s+t-1)}{\Gamma(-s+t)}-\frac{1}{t^{1-2s} } \ , 
\end{align}
which subtracts out the $\frac{1}{t^{1-2s}}$ tail from the ratio of the gamma functions. Under the conditions $t\ge 1$ and $s=c+iy$ with $0<c<1/2$, $y\in R$,  the remainder function satisfies the following inequality
\begin{align}
 &\bigg|r(s,t)\bigg| \le A(c)\bigg|\frac{1-2s}{t^{2-2s}} \bigg| \ ,
\end{align}
where $0<A(c)=\frac{\frac{\Gamma(c)}{\Gamma(1-c)}-1}{1-2c}<\infty$. Given the above, one has the natural decompositions
\begin{align}
&\hat q_1^{(l)}(t,\lambda)=\hat q_1^{(l),\frac{1}{t^{1-2s}}}(t,\lambda)+\hat q_1^{(l),r}(t,\lambda) \ , \\ 
&{\cal F}_1^{(l)}(z^2m^2,t,\lambda)={\cal F}_1^{(l),\frac{1}{t^{1-2s}}}(z^2m^2,t,\lambda)+{\cal F}_1^{(l),r}(z^2m^2,t,\lambda) \ .
\end{align}
For the remainder terms, one simply bound using absolute values 
\begin{align}
|\hat q_1^{(l),r}(t,\lambda)| \le \frac{A(c)}{4\pi l!} \bigg|\frac{\Gamma(-c-l)\Gamma(2l+2+2c)}{t^{2-2c}\Gamma(1+2l+c)}\bigg| \int_{-\infty}^{\infty} \frac{dy}{2\pi} \bigg|(2c+2iy-1)\Gamma(1-2c-2iy)\bigg| \ .
\end{align}
Clearly, the remaining integral converges and is bounded by $0<B(c)<\infty$. Thus one has for arbitrary $N\gg1$, $M\gg 1$  (using the shifted $t$)
\begin{align}
&\bigg|\int_{1}^{N} dt \sum_{l=0}^{M}\left(\frac{z^2m^2}{4}\right)^l \hat q_1^{(l),r}(t,\lambda) \bigg| \nonumber \\ 
\le &\sum_{l=0}^{\infty}\int_{1}^{\infty}dt \left(\frac{z^2m^2}{4}\right)^l \bigg|\hat q_1^{(l),r}(t,\lambda)\bigg| \nonumber \\ 
\le &\sum_{l=0}^{\infty}\left(\frac{z^2m^2}{4}\right)^l \frac{A(c)B(c)}{4\pi l!} \bigg|\frac{\Gamma(-c-l)\Gamma(2l+2+2c)}{(1-2c)\Gamma(1+2l+c)}\bigg|<\infty \ .
\end{align}
This implies that the power expansion of the  remainder part for $q_1^{(l)}$ converges absolutely. Similarly, one can show that the twist expansion in terms of the remainder part for ${\cal F}_1^{(l)}$ converges absolutely.  

One then consider the $\frac{1}{t^{1-2s}}$ parts which contain the large $t$ tail. For $q_1$ one has 
\begin{align}
\hat q_1^{(l),\frac{1}{t^{1-2s}}}(t,\lambda)=\frac{(-1)^l}{4\pi l!}\int_{c-i\infty}^{c+i\infty}\frac{ds}{2\pi i}&\frac{\Gamma(-l-s)\Gamma(2l+2+2s)\Gamma(1-2s)}{t^{1-2s}}\nonumber \\ & \times \,_1\tilde F_1(2+2l+2s,1+2l+s,-\lambda) \ .
\end{align}
Now, to isolate the large $t$ tail, one can simply shift the contour to $c \rightarrow c-\frac{1}{2}$ by picking up the pole at $s=0$. This leads exactly to the large $t$ tail that will be removed by the subtraction terms in the scale-dependent version Eq.~(\ref{eq:q1lmu}), on top of another Mellin-integral which is now bonded by $\frac{1}{t^{2-2c}}$. Still using previous bounds (which hold for $-\frac{1}{2}<c<0$ as well), one can show in a straightforward manner that the $t$-integral over the sum after subtracting the large $t$ tails is dominated by absolute convergent sums. More precisely, using the un-shifted $t$, one has
\begin{align}
&\bigg|\hat q_1^{(l),\frac{1}{t^{1-2s}}}(t,\lambda)+\frac{\Gamma(2l+2)}{4\pi (l!)^2}\,_1\tilde F_1(2+2l,1+2l,-\lambda)\frac{1-\mu^t}{t} \bigg| \nonumber \\ 
\le &\frac{\Gamma(2l+2)}{4\pi (l!)^2\Gamma(1+2l)} \bigg|\frac{1-\mu^t}{t}-\frac{1}{t+1}\bigg|+\frac{B(c')}{4\pi l!(t+1)^{1-2c'}}\bigg|\frac{\Gamma(-l-c')\Gamma(2l+2+2c')}{\Gamma(1+2l+c')}\bigg| \ ,
\end{align}
with $-\frac{1}{2}<c'<0$ and $0<B(c')<\infty$, for example one can simply chose $c'=-\frac{1}{5}$. Multiplying $\left(\frac{z^2m^2}{4}\right)^l$, integrating over $t$ and summing over $l$ leads to the desired result.  The case of ${\cal F}_1^{(l),\frac{1}{t^{2s-1}}}$ can be treated similarly.  This finally shows that one has (in the un-shifted $t$)
\begin{align}
&\bigg|\int_{0}^{N} dt \sum_{l=0}^{M}\left(\frac{z^2m^2}{4}\right)^l \bigg(q_1^{(l)}(t,\lambda,\mu)+{\cal F}_{1}^{(l)}(z^2m^2,t,\lambda,\mu)\bigg)\bigg| \nonumber \\ 
\le &\sum_{l=0}^{\infty}\left(\frac{z^2m^2}{4}\right)^l\int_{0}^{\infty} dt \bigg|q_1^{(l)}(t,\lambda,\mu)\bigg|+\sum_{l=0}^{\infty}\left(\frac{z^2m^2}{4}\right)^l\int_{0}^{\infty} dt \bigg|{\cal F}_{1}^{(l)}(z^2m^2,t,\lambda,\mu)\bigg|\nonumber \\<&\infty \ .
\end{align}
As a result, the power expansion for $F_1$ in terms of $q_1^{(l)}$ and ${\cal F}_{1}^{(l)}$ converges absolutely for any $z^2>0$. Moreover, the $t$ integral and the sum can be exchanged freely.  

\subsection{Expansion in $q_1^{(l)}$, ${\cal H}_1^{l,p}$ and $q_1^{l,p}$}
We then show that the full twist expansion of $F_1$ in terms of the $q^{(l)}_1$, the ${\cal H}_1^{l,p}(\alpha(z),\lambda,\mu)$ given by Eq.~(\ref{eq:H1lpmu}) and the $q_1^{l,p}(\lambda,\mu)$ given by Eq.~(\ref{eq:q1lmu}) is absolutely convergent. For this purpose, it is sufficient to consider $l\ge 1$ and $p\ge 1$, since the special cases $l=0$ or $p=0$ pose no danger to absolute convergence and can be treated separately. We need to show that after taking the absolute value of individual terms, integrating over $t$ term by term, and then summing over $p$, the sum over $l$ is still convergent. We chose the integration path as $(1+i\epsilon)t$ with $0<t<\infty$ and $\epsilon=\frac{1}{10}$. We consider $0<t<1$ and $t \ge 1$ parts separately. We first consider the $t\ge 1$ part, which is easier to deal with. We need the inequalities under the above conditions
\begin{align}
&\bigg|\Gamma(2l+1+p+(1+i\epsilon)t)\bigg|\ge \Gamma(2l)\bigg|\Gamma(1+p+(1+i\epsilon)t)\bigg| \ , \\
&\bigg|\frac{\Gamma(2(1+i\epsilon)t+2p+1)}{\Gamma(1+2(1+i\epsilon)t+p)\Gamma(1+p+(1+i\epsilon)t)}\bigg| \le \frac{\Gamma(2p+1)}{(p!)^2} \  , \\
&\bigg|\Gamma((1+i\epsilon)t+p+1)\bigg| \ge \bigg|\Gamma((1+i\epsilon)t+1)\bigg|\Gamma(p) \ , \\ 
&(l+p)!\ge l!p! \ .
\end{align}
The above lead to $t\ge 1$, $0<z^2<\rho<\infty$
\begin{align}
\bigg|{\cal H}_1^{l,p}(t(1+i\epsilon ),\alpha(z),\lambda)\bigg| \le \left(\frac{m^2\rho^2}{4}\right)^t\frac{\Gamma(2l+2)\Gamma(2p+1)}{4\pi\Gamma(2l)l! (p!)^2 p! }\bigg|\Gamma(-l-p-(1+i\epsilon)t)\bigg| \ , \\
\bigg|q_1^{l,p}(t(1+i\epsilon ),\lambda)\bigg| \le \frac{\Gamma(2l+2)\Gamma(2p+1)}{4\pi (2l)! (l!)^2 \Gamma(p) p!}\bigg|\frac{\Gamma((1+i\epsilon)t-p)}{\Gamma((1+i\epsilon)t+1)}\bigg| \ .
\end{align}
Now, under the above conditions, one has
\begin{align}
&\bigg|\Gamma(-l-p-(1+i\epsilon)t)\bigg|=\bigg|\frac{\pi}{\sin \pi(1+i\epsilon)t \Gamma(l+p+1+(1+i\epsilon)t)}\bigg|\nonumber \\ 
&\le \frac{1}{l!p!}\bigg|\frac{\pi}{\sin \pi(1+i\epsilon)t}\bigg| \bigg|\frac{1}{\Gamma((1+i\epsilon)t)}\bigg| \ ,
\end{align}
and 
\begin{align}
\bigg|\frac{\Gamma((1+i\epsilon)t-p)}{\Gamma((1+i\epsilon)t+1)}\bigg| \le \frac{1}{\epsilon^{p+1} t^{p+1}} \ ,
\end{align}
thus one has
\begin{align}
&\sum_{p=1}^{\infty}\left(\frac{z^2m^2}{4}\right)^p\int_{1}^{\infty} dt \bigg|{\cal H}_1^{l,p}(t(1+i\epsilon ),\alpha(z),\lambda)\bigg| \le \frac{B(\epsilon,\rho)\Gamma(2l+2)}{4\pi \Gamma(2l)(l!)^2} \ , \\
&\sum_{p=1}^{\infty}\left(\frac{z^2m^2}{4}\right)^p\int_{1}^{\infty} dt \bigg|{q}_1^{l,p}(t(1+i\epsilon ),\lambda)\bigg| \le \frac{A(\epsilon,\rho)\Gamma(2l+2)}{4\pi (2l)!(l!)^2} \ ,
\end{align}
with $0<A(\epsilon,\rho)<\infty$ and $0<B(\epsilon,\rho)<\infty$. The subtraction terms in the $t\ge 1$ region depends on $t$ only through the harmless combination $\frac{\mu^t}{t}$ and can be bounded similarly. This leads to the desired results for $t\ge 1$ part. 

One now moves to the trickier $0<t<1$ part. We now introduce the notation
\begin{align}
&{\cal K}(l,p,\omega,\lambda)\nonumber \\ 
&=\left(\frac{z^2m^2}{4}\right)^{\omega}\frac{\Gamma(2\omega+2p+1)}{\Gamma(2\omega+p+1)\Gamma(1+2l+p+\omega)}\,_1F_1(2+2l,1+2l+p+\omega,-\lambda) \ .
\end{align}
We now need to bound this function and its {\it derivative} with respect to $\omega=(1+i\epsilon)t$  in $ 0< t<1$. For the function itself, it is simple: one has ($l\ge 1$, $p\ge 1$, $\lambda>0$, $0<c_0<\infty$)
\begin{align}
\bigg|{\cal K}(l,p,(1+i\epsilon)t,\lambda)\bigg| \le \frac{c_0\Gamma(2p+1)}{(p!)^2\Gamma(2l)} \ .
\end{align}
For the derivative, notice that the absolute value of the digamma functions $\bigg|\frac{\Gamma'(\omega)}{\Gamma(\omega)}\bigg|$ in the current case can simply be bounded by $\ln(2p+1)+c_1$, $\ln(p+1)+c_2$, $\ln(2l+p+1)+c_3$. So we need only to bound the derivative of the $_1F_1$. Now since $2+2l>2$ and ${\rm Re}(1+2l+p+\omega-2-2l)>0$, one always has 
\begin{align}
\,_1F_1(2+2l,1+2l+p+\omega,-\lambda)=\frac{\Gamma(1+2l+p+\omega)}{\Gamma(2+2l)\Gamma(p+\omega-1)}\int_0^1 d\alpha e^{-\alpha \lambda}\alpha^{2l+1}(1-\alpha)^{p+\omega-2} \ .
\end{align}
Thus for $p>1$, $ 0<t<1$ and $l\ge 1$, using  \begin{align}
&\int_{0}^1 d\alpha \alpha^{2l+1}(1-\alpha)^{p+t-2}\ln \frac{1}{1-\alpha}\nonumber \\ 
&=-\frac{\Gamma (2 l+2) \Gamma (p+t-1) \left(\psi(p+t-1)-\psi(2l+1+p+t)\right)}{\Gamma (2 l+p+t+1)} \ , \nonumber \\ 
\end{align}
and
\begin{align}
\bigg|\frac{\Gamma(1+2l+p+t(1+i\epsilon))}{\Gamma(p+t(1+i\epsilon)-1)}\frac{\Gamma(p+t-1)}{\Gamma(2l+p+1+t)}\bigg| \le A<\infty \ ,
\end{align}
one can bound the derivative of $_1F_1$ by logarithms 
\begin{align}
\bigg|\frac{d}{d\omega} \,_1F_1(2+2l,1+2l+p+\omega,-\lambda)\bigg| \le (c_4\ln l+c_5\ln p+c_6) \ ,
\end{align}
with $0<c_4,c_5,c_6<\infty$. For the special case $p=1$, one can write
\begin{align}
-\frac{1}{\Gamma(\omega+1)}\int_{0}^1 d\alpha e^{-\alpha \lambda}\alpha^{2l+1}\frac{d}{d\alpha}(1-\alpha)^{\omega} \ ,
\end{align}
and partial integrate to proceed. At worst, one obtains ${\cal O}(l^2)$, ${\cal O}(p^2)$, ${\cal O}(\lambda)$. Given the above, one has
\begin{align}
\bigg|\frac{d}{d\omega}{\cal K}(l,p,\omega,\lambda)\bigg|_{\omega=(1+i\epsilon)t} \le \frac{\Gamma(2p+1)}{(p!)^2\Gamma(2l)}\bigg(a_1l^2+a_2p^2+a_3\bigg) \ ,
\end{align}
where all constants are positive finite numbers. Now, to include the subtraction term, one simply writes the subtraction as 
\begin{align}
&(-1)^{l+p}{\cal H}_1^{l,p}((1+i\epsilon)t,\alpha(z),\lambda,\mu)\nonumber \\ 
&=\frac{\Gamma(2l+2)}{4\pi l!p!}{\cal K}(l,p,(1+i\epsilon)t,\lambda)\bigg(\Gamma(-l-p-t(1+i\epsilon))+\frac{(-1)^{l+p}\mu^t}{(l+p)!(1+i\epsilon)t}\bigg) \nonumber \\ 
&- \frac{\Gamma(2l+2)\mu^t}{4\pi l!p!}\int_{0}^1 ds \frac{d}{d\omega}{\cal K}(l,p,\omega,\lambda)\bigg|_{\omega=(1+i\epsilon)st}\frac{(-1)^{l+p}}{(l+p)!} \ .
\end{align}
The integral in the third line can be bounded without trouble now. Using 
\begin{align}
\Gamma(-\omega-l-p)=\frac{(-1)^{l+p}\Gamma(-\omega)}{(\omega+1)...(\omega+l+p)} \ ,
\end{align}
one can bound the  bracket in the second line by $\frac{a_4\ln (l+p)}{(l+p)!}$. Thus, one finally obtains 
\begin{align}
\int_{0}^1 dt \bigg|{\cal H}_1^{l,p}((1+i\epsilon)t,\alpha(z),\lambda,\mu)\bigg|\le \frac{\Gamma(2l+2)\Gamma(2p+1)}{4\pi \Gamma(2l) (p!)^2(l!)^2(p!)^2}\bigg(a_1'l^2+a_2'p^2\bigg) \ ,
\end{align}
with $0<a_1'<\infty$, $0<a_2'<\infty$. Clearly the sum over $l\ge 1$, $p\ge 1$ is absolutely convergent after multiplying with $\left(\frac{z^2m^2}{4}\right)^{l+p}$. Combining with the case of $t\ge 1$ one obtains for any $0<z^2<\infty$ and $0\le \lambda<\infty$
\begin{align}
\sum_{l\ge 1 }\sum_{p\ge 1}\left(\frac{z^2m^2}{4}\right)^{l+p}\int_{0}^\infty dt \bigg|{\cal H}_1^{l,p}((1+i\epsilon)t,\alpha(z),\lambda,\mu)\bigg|<\infty \ .
\end{align}
The case of $q_1^{l,p}$ can be treated similarly. Combining with the separate cases $p=0$ or $l=0$ one finally shows that the full twist expansion is also absolutely convergent.

\section{Threshold asymptotics of the structure function}\label{sec:threshold}
In this section, we investigate another important kinematic limit, the {\it threshold limit} defined as $\lambda \rightarrow +\infty$~\cite{Liu:2023onm}. This limit is closely related to the DIS-like threshold limit 
$x_B \rightarrow 1$ at fixed $Q^2$, for which the partoinc version seems to support universal massless asymptotics such as the ``Sudakov'' type asymptotics in QCD~\cite{Sen:1981sd, Mueller:1984vh,Korchemsky:1988pn,Korchemsky:1988hd}, but the full theory does not. The point is that in the standard formulation of the threshold limit, as $x_B \rightarrow 1$, no matter how large the $Q^2$, the invariant mass $M^2+Q^2(\frac{1}{x_B}-1)$ of the final state will ultimately become too soft to be controllable by the partonic descriptions. In particular, the $Q^2 \rightarrow \infty$ twist expansion and the $x_B \rightarrow 1$ threshold expansion will ultimately compete with each other. 

On the other hand, in the coordinate space formulation of the threshold limit, the lower bound for the involved hard scale is $-z^2$ and remains perturbative as far as $-z^2$ is small, no matter how large the $\lambda$ is. This leads to the conjecture in Ref.~\cite{Liu:2023onm} that not only the leading-twist coordinate space coefficient function in the threshold limit allows ``genuine-renormalon free'' threshold factorization (in QCD), but the twist and threshold expansions at the level of the coordinate-space hadronic matrix elements are also mutually consistent. This serves as an important motivation for the calculations in this section, namely, to validate or invalidate the above statements. In Ref.~\cite{Liu:2023onm}, we also found that the ``resurgent relation''~\cite{Basso:2006nk} between the threshold and the ``Regge'' asymptotics has a nice representation in the coordinate space, which will also be demonstrated explicitly in this section.   

To perform the threshold expansion,  one introduces one more layer of Mellin representation in $\lambda$:
\begin{align}
F_1(u,s,\lambda)=\frac{1}{4\pi}\Gamma(u)\Gamma(u-s) \int\frac{du_1}{2\pi i}\frac{\Gamma(-u_1)\Gamma(2+2s-2u+u_1)}{\Gamma(1+s-2u+u_1)}\lambda^{u_1} \ .
\end{align}
Picking up residues at $u_1=-2-2s+2u-k$, one obtains
\begin{align}
&F_1(u,s,\lambda)\nonumber \\ 
&\sim \frac{1}{4\pi}\sum_{k=0}^{\infty}\Gamma(u)\Gamma(u-s)\frac{(-1)^k \Gamma (k+2 s-2 u+2)}{k! \Gamma (-k-s-1) }\lambda^{-k-2-2s+2u}\left(\frac{z^2m^2}{4}\right)^{-u} \ .
\end{align}
Now, by picking up poles in $u$, there are two series. One is 
\begin{align}\label{eq:q1thresoft}
&F_1^{s}(z^2m^2,s,\lambda)\nonumber \\ 
&\sim \frac{1}{4\pi}\sum_{k=0}^{\infty}\sum_{l=0}^{\infty}\Gamma(-l-s)\frac{(-1)^{k+l}}{k!l!}\frac{\Gamma(k+2l+2s+2)}{\Gamma(-k-s-1)}\lambda^{-k-2l-2-2s}\left(\frac{z^2m^2}{4}\right)^l \ .
\end{align}
Another series is 
\begin{align}
F_1^{h}(z^2m^2,s,\lambda)\sim \frac{1}{4\pi}\sum_{k=0}^{\infty}\sum_{l=0}^{\infty}\Gamma(s-l)\frac{(-1)^{k+l}}{k!l!}\frac{\Gamma(k+2l+2)}{\Gamma(-k-s-1)}\lambda^{-k-2l-2}\left(\frac{z^2m^2}{4}\right)^{-s+l} \ .
\end{align}
The ``leading-power'' large $\lambda$ terms correspond to $k=l=0$ (this will be justified in a moment), namely 
\begin{align}
F_1(z^2m^2,s,\lambda)\sim \frac{1}{4\pi}\frac{\Gamma(-s)\Gamma(2s+2)}{\Gamma(-s-1)}\lambda^{-2-2s}+\frac{1}{4\pi}\frac{\Gamma(s)}{\Gamma(-s-1)}\lambda^{-2}\left(\frac{z^2m^2}{4}\right)^{-s} \ .
\end{align}
The second term leads to
\begin{align}
\frac{1}{\lambda^2}{\cal F}^{l=k=0}_1(z^2m^2)=\frac{1}{4\pi\lambda^2}\int_{0}^{\infty} dt \int_{c-i\infty}^{c+i\infty} \frac{ds}{2\pi i}\frac{\Gamma (1-2 s) \Gamma (s) \Gamma (s+t)}{\Gamma (-s-1) \Gamma (-s+t+1)}\left(\frac{m^2 z^2}{4}\right)^{-s} \ .
\end{align}
It plays the role of the ``threshold soft factor'' in the current case. At small $z$, it simply reduces to the large $\lambda$ expansion of the hard kernel. On the other hand, the first term contributes to 
\begin{align}
q_{1}^{l=k=0}(\lambda)=-\frac{1}{4\pi\lambda^2}\int_{0}^{\infty} dt \int_{c-i\infty}^{c+i\infty} \frac{ds}{2\pi i}\frac{(s+1)\Gamma (1-2 s)  \Gamma(2s+2)\Gamma (s+t)}{\Gamma (-s+t+1)}\left(\lambda^2\right)^{-s} \ .
\end{align}
One needs to understand the large $\lambda$ decay of this type of term, which is required in order to show that it is indeed $k=l=0$ in $F^{s}_1$ that is leading in the large $\lambda$ limit.

\subsection{Analytic continuation and values at positive integers of a crucial function}\label{sec:anay}
In order to understand the large $\lambda$ behavior for functions such as $q_1^{l=k=0}(\lambda)$, we need to introduce the analytic continuation of a crucial function to the right-half plane. We start from the region $0<{\Re}(s)<\frac{1}{2}$. The function $f(s)$ is defined as
\begin{align}
f(s)=-\frac{1}{2s}+\int_{1}^{\infty}  \bigg(\frac{\Gamma(s+t-1)}{\Gamma(t-s)}-\frac{1}{t^{1-2s}}\bigg)dt \ .
\end{align}
Clearly, the integral converges absolutely in this region and defines an analytic function in $s$. We can continue further to the right by adding and subtracting. Up to $0<{\Re}(s)<\frac{3}{2}$, one needs
\begin{align}
&f(s)=\frac{(2 s-1) (s^2-6s+6)}{12 s}\nonumber \\ 
&+\int_{1}^{\infty}  \bigg(\frac{\Gamma(s+t-1)}{\Gamma(t-s)}-\left(\frac{1}{t^{1-2s}}+\frac{1-2 s}{t^{2-2s}}-\frac{(s-6) (s-1) (2 s-1)}{6 t^{3-2s}}\right) \bigg)dt \ .
\end{align}
Now, one would like to calculate the values (or singularities) at $s=\frac{1}{2}$ and $s=1$. Generically, one needs to perform the integral numerically, but for these two values, the integrands all vanish. Thus one obtains directly $f\left(\frac{1}{2}\right)=0$ and $f\left(1\right)=\frac{1}{12}$. One can continue this way further by subtracting out more tails. For any integer or half-integer $s$, once one reaches the convergence region of the subtracted integral, the integrand vanishes. This is due to the fact that for these $s$, the un-subtracted integrands $\frac{\Gamma(s+t-1)}{\Gamma(t-s)}$ are finite order polynomials in $t$. Thus, values or singularities of the $f(s)$ at integer and half-integer values can all be calculated in this manner through a finite number of subtractions. 

In fact, one has the asymptotic expansions in the $t\rightarrow \infty$ limit
\begin{align}
\frac{\Gamma(t+s-1)}{\Gamma(t-s)}\sim \sum_{k=0}^{\infty}\binom{2s-1}{k}\frac{B_k^{(2s)}(s-1)}{t^{1-2s+k}} \ ,
\end{align}
where the $B_k^{(\alpha)}(z)$ are the generalized Bernoulli polynomials.  From this, one obtains for integer or half-integer $s$
\begin{align}
f(s)=-\sum_{k=0}^{2s}\frac{\Gamma(2s)}{k!\Gamma(2s+1-k)}B_k^{(2s)}(s-1) \ .
\end{align}
Notice that they are all finite. On the other hand, since all the subtracted integrals in the convergence regions are also finite, singularities of the function $f(s)$ can only be generated by the tails and located at integer or half-integer values. The finiteness of $f(s)$ at integer and half-integer values then implies that {\it $f(s)$ is analytic in the whole right-half plane}. 

Below, we provide compact expressions for $f(s)$ at integer or half-integer values. We first consider half-integer cases. For $2s=2p+1$, explicitly one has
\begin{align}
f\left(\frac{2p+1}{2}\right)=-\sum_{k=0}^{2p+1}\frac{\Gamma(2p+1)}{k!\Gamma(2p+2-k)}B_k^{(2p+1)}\left(\frac{2p-1}{2}\right) \ .
\end{align}
To evaluate this sum, one needs to use the following definitions for the generalized Bernoulli polynomials
\begin{align}
B_k^{l}(z)=\frac{k!}{(2\pi i)}\oint \frac{dw}{w^{k+1}}\frac{w^le^{wz}}{(e^{w}-1)^l} \ ,
\end{align}
where one integrates over a small circle with a radius smaller than $2\pi$ around the origin. Given the above, one obtains, after summing over $k$, the compact expression in terms of the incomplete gamma functions as
\begin{align}
f\left(\frac{2p+1}{2}\right)=-\frac{1}{2p+1}\oint \frac{dw}{2\pi i w} \frac{ \Gamma (2 p+2,w)}{(e^{\frac{w}{2}}-e^{-\frac{w}{2}})^{2p+1}} \ .
\end{align}
Now, the incomplete gamma function's expansion at $w=0$ is of the form $\Gamma(2p+2)-\frac{w^{2p+2}}{2p+2}-..$, while the denominator's expansion is of the form $\frac{1}{w^{2p+1}}(1+a_1w^2+...)$. Thus, no residue is produced in the small $w$ expansion. Therefore, we have shown that 
\begin{align}
f\left(\frac{2p+1}{2}\right) \equiv 0 \ , \forall p\in Z_{\ge 0} \ .
\end{align}
We then consider the integer cases. Using the same method, one has
\begin{align}
f(p)=-\frac{1}{2p}\oint \frac{dw}{2\pi i w} \frac{ \Gamma (2 p+1,w)}{(e^{\frac{w}{2}}-e^{-\frac{w}{2}})^{2p}}\ , \forall p\in Z_{>0} \ .
\end{align}
Unlike the case of half-integers, it has a non-vanishing residue at $w=0$. Moreover, since the incomplete gamma function subtracted $\Gamma(2p+1)$ starts from $w^{2p+1}$, one has
\begin{align}
f(p)=-\frac{\Gamma(2p+1)}{2p}\oint \frac{dw}{2\pi i w} \frac{ 1}{(e^{\frac{w}{2}}-e^{-\frac{w}{2}})^{2p}} \ .
\end{align}
From this, one simply picks the constant term of the expansion for the $(2\sinh \frac{w}{2})^{-2p}$ at $w=0$ to evaluate $f(p)$. In particular, the values of $f(s)$ at $s=1,2,3,4$ are calculated as
\begin{align}
\{f(p)\}_{1 \le p\le 4}=\bigg\{\frac{1}{12},-\frac{11}{120},\frac{191}{504},-\frac{2497}{720}\bigg\} \ .
\end{align}
To proceed further, one shifts the contour to $w=\pm i\pi+y$, $y\in R$ (this is actually a complex Lorentz-transform). This leads to
\begin{align}
f(p)=(-1)^{p-1}\frac{\Gamma(2p+1)}{2^{2p+1}p}\int_{-\infty}^{\infty}\frac{dy}{y^2+\pi^2}\frac{1}{ \cosh^{2p} \frac{y}{2}} \ ,
\end{align}
with the sharp large $p$ asymptotics. From this, it is then straightforward to combine with residues of the $\Gamma(1-2s)$ to establish the spectral representation for the effective coupling $g^2(k^2)$
\begin{align}
g^2(-k^2)=\frac{\pi}{N}\int_{-\infty}^{\infty}\frac{dy}{y^2+\pi^2}\frac{4m^2\cosh^2 \frac{y}{2}}{-k^2+4m^2\cosh^2 \frac{y}{2}} \ ,
\end{align}
with the correct location of the branch cut $(4m^2,\infty)$.

\subsection{Threshold expansions for $F_1$, $F_2$ and $F_3$}

Given the knowledge of the analytic continuation of the function $f(s)$ and its values at positive integers, we are able to perform the threshold expansion for all the obtained functions.

To obtain the large $\lambda$ expansion, we use the trick as follows. We first make the change of variable $t\rightarrow t+1$ and split the $t$-dependent gamma function ratio into the remainder and the $\frac{1}{t^{1-2s}}$ tail:
\begin{align}
q_1^{l=k=0}(\lambda)=-\frac{1}{4\pi\lambda^2}\int_{1}^{\infty} dt \int_{c-i\infty}^{c+i\infty} \frac{ds}{2\pi i}(s+1)\Gamma (1-2 s)  \Gamma(2s+2)r(s,t)\left(\lambda^2\right)^{-s} \nonumber \\ 
-\frac{1}{4\pi\lambda^2}\int_{1}^{\infty} dt \int_{c-i\infty}^{c+i\infty} \frac{ds}{2\pi i}(s+1)\Gamma (1-2 s)  \Gamma(2s+2)\frac{1}{t^{1-2s}}\left(\lambda^2\right)^{-s} \ .
\end{align}
Notice that the remainder function is defined in Eq.~(\ref{eq:remainder}). Now, due to the removal of the large $t$ tail, the integral in the first line converges absolutely, and one can exchange the order of the $s,t$ integrals. In the second line, on the other hand, one can first shift the $s$-contour to $-\frac{1}{2}<c'<0$ to reach the domain of absolute convergence and then exchange the $s$, $t$ integrals. One then would like to shift back to $0<c<\frac{1}{2}$ and combine with the $\int_1^{\infty} dt r(s,t)$ to form the $f(s)$. However, the $t$ integral of the tail $\frac{1}{t^{1-2s}}$ is $-\frac{1}{2s}$ and contains an extra pole at $s=0$. When shifting the $s$ contour back, this pole must be included as well. This finally leads to
\begin{align}
q_1^{l=k=0}(\lambda)=-\frac{1}{8\pi \lambda^2}-\frac{1}{4\pi\lambda^2} \int_{c-i\infty}^{c+i\infty} \frac{ds}{2\pi i}(s+1)\Gamma (1-2 s)  \Gamma(2s+2)f(s)\left(\lambda^2\right)^{-s} \ .
\end{align}
Notice that the first term is due to the $s=0$ pole of the $-\frac{1}{2s}$ and corresponds to the {\it leading} large $\lambda$ asymptotics.

One now shifts the $s$-contour further to the right and combines the values of $f(s)$ at positive integers (this explains why we investigate this function in the previous subsection) with the residues of $(s+1)\Gamma(1-2s)\Gamma(2s+2)$  to obtain the threshold expansion of $q_1^{l=k=0}$ to arbitrary threshold powers:
\begin{align}
q_1^{l=k=0}(\lambda)=-\frac{1}{4\pi \lambda^2}\bigg(\frac{1}{2}-\frac{1}{2\lambda^2}+\frac{11}{4\lambda^4}-\frac{191}{6\lambda^6}+{\cal O}\left(\frac{1}{\lambda^8}\right)\bigg) \ .
\end{align}
In particular, the $-\frac{1}{8\pi \lambda^2}$ is nothing but the leading large $\lambda$ asymptotics of $q_1(\lambda)$ . To obtain the full expansion for $q_1$ beyond the $\frac{1}{\lambda^2}$ tail, one should also include the $k \ge 1$ terms in Eq.~(\ref{eq:q1thresoft}), which can be expanded similarly. On the other hand, for higher $k$ and $l$, the corresponding large $\lambda$ asymptotics is always $\frac{1}{\lambda^{k+2l+2}}$ times at worst $\lambda^0$. Since $l$ also counts the twist, this implies that the threshold expansion and the small $z$ expansion are mutually consistent. In particular, {\it higher-twist non-perturbative functions decay faster in the $\lambda \rightarrow +\infty$ limit}.

Similarly, for $F_2(u,s,\lambda)$ one has 
\begin{align}
F_2(u,s,\lambda)=-\frac{s}{\pi}\Gamma(u)\Gamma(u-s+1) \int\frac{du_1}{2\pi i}\frac{\Gamma(-u_1)\Gamma(2s-2u+u_1)}{\Gamma(1+s-2u+u_1)}\lambda^{u_1} \ .
\end{align}
This leads to
\begin{align}
&F_2(u,s,\lambda) \nonumber \\
&\sim -\frac{s}{\pi}\sum_{k=0}^{\infty}\Gamma(u)\Gamma(u-s+1)\frac{(-1)^k\Gamma(2s-2u+k)}{k!\Gamma(-k-s+1)}\lambda^{-k-2s+2u} \ .
\end{align}
The $u=0$ pole with $k=0$ leads to
\begin{align}
q_2^{l=k=0}(\lambda)=-\frac{1}{\pi}\int_{0}^{\infty} dt \int_{c-i\infty}^{c+i\infty} \frac{ds}{2\pi i}\frac{s\Gamma (1-2 s)  \Gamma(2s)\Gamma (s+t)}{\Gamma (-s+t+1)}\left(\lambda^2\right)^{-s} \ .
\end{align}
Now, the leading asymptotics up to $\frac{1}{\lambda^2}$ reads $-\frac{1}{4\pi}+\frac{1}{24\pi \lambda^2}$. To get this, one also needs the fact that for $(l,k)=(0,1)$ and $(l,k)=(0,2)$, the pole at $s=0$ is cancelled and the asymptotics starts at $\frac{1}{\lambda^3}$ and $\frac{1}{\lambda^4}$, respectively.  Notice that for $u=-1$ and $u=s-1$ one also has $\frac{1}{\lambda^2}$ contributions from $k=0$. For the former, one has
\begin{align}
&\frac{z^2m^2}{4}q^{l=1,k=0}_2(\lambda)\nonumber \\ 
&=\left(\frac{z^2m^2}{4\lambda^2}\right)\int_{0}^{\infty} dt \int_{c-i\infty}^{c+i\infty} \frac{ds}{2\pi i}\frac{s}{\pi}\frac{\Gamma(-s)\Gamma(2s+2)\Gamma(1-2s)\Gamma(s+t)}{\Gamma(-s+1)\Gamma(-s+t+1)}(\lambda^2)^{-s} \ .
\end{align}
The leading asymptotics reads $-\frac{z^2m^2}{8\pi \lambda^2}$.  For the latter, one has
\begin{align}
&\frac{z^2m^2}{4\lambda^2}{\cal F}_2^{l=1,k=0}(z^2m^2)\nonumber \\ 
&=-\left(\frac{z^2m^2}{4\lambda^2}\right)\int_{0}^{\infty} dt \int_{c-i\infty}^{c+i\infty} \frac{ds}{2\pi i}\frac{s}{\pi}\frac{\Gamma(s-1)\Gamma(1-2s)\Gamma(s+t)}{\Gamma(-s+1)\Gamma(-s+t+1)}\left(\frac{z^2m^2}{4}\right)^{-s} \ .
\end{align}
In particular, by performing the small $z$ expansion on the above, one obtains the large $\lambda$ expansion for the hard kernel ${\cal H}_2^{1,0}$.

The case of $F_3$ can be analyzed as well. Due to the overall $\lambda$, one needs both $l=0,k=0$ and $l=0,k=1$ to reach $\frac{1}{\lambda^2}$.  For the soft part, one has
\begin{align}
q_3^{l=k=0}(\lambda)=\frac{1}{2\pi\lambda}\int_{0}^{\infty} dt \int_{c-i\infty}^{c+i\infty} \frac{ds}{2\pi i}\frac{\Gamma (1-2 s)  \Gamma(2s+2)\Gamma (s+t)}{\Gamma (-s+t+1)}\left(\lambda^2\right)^{-s} \ .
\end{align}
The leading large $\lambda$ asymptotics reads $\frac{1}{4\pi \lambda}$. The $k=1$ term reads
\begin{align}
q_3^{l=0,k=1}(\lambda)=\frac{1}{2\pi\lambda^2}\int_{0}^{\infty} dt \int_{c-i\infty}^{c+i\infty} \frac{ds}{2\pi i}\frac{(s+1)\Gamma (1-2 s)  \Gamma(2s+3)\Gamma (s+t)}{\Gamma (-s+t+1)}\left(\lambda^2\right)^{-s} \ .
\end{align}
The large $\lambda$ tail reads $\frac{1}{2\pi \lambda^2}$. From the above one obtains $q_3^{(0)}(\lambda) \rightarrow \frac{1}{4\pi \lambda}+\frac{1}{2\pi \lambda^2}+{\cal O}\left(\frac{1}{\lambda^3}\right)$. 
For the ``hard part'', one has
\begin{align}
\frac{1}{\lambda}{\cal F}_3^{l=k=0}(z^2m^2)=\frac{1}{2\pi\lambda}\int_{0}^{\infty} dt \int_{c-i\infty}^{c+i\infty} \frac{ds}{2\pi i}\frac{\Gamma (1-2 s) \Gamma (s) \Gamma (s+t)}{\Gamma (-s) \Gamma (-s+t+1)}\left(\frac{m^2 z^2}{4}\right)^{-s} \ ,
\end{align}
as well as
\begin{align}
\frac{1}{\lambda^2}{\cal F}_3^{l=0,k=1}(z^2m^2)=-\frac{1}{\pi\lambda^2}\int_{0}^{\infty} dt \int_{c-i\infty}^{c+i\infty} \frac{ds}{2\pi i}\frac{\Gamma (1-2 s) \Gamma (s) \Gamma (s+t)}{\Gamma (-s-1) \Gamma (-s+t+1)}\left(\frac{m^2 z^2}{4}\right)^{-s} \ .
\end{align}
Notice only $l=0$ contributes to the $\frac{1}{\lambda}$ and $\frac{1}{\lambda^2}$ tails in the $F_1$ and $F_3$ series. The above exhausted all the contributions up to $\frac{1}{\lambda^2}$ for the structure function in the threshold limit: 
\begin{align}\label{eq:Ethreshold}
&\frac{Ne^{-\lambda}}{2\pi}{\cal E}(z^2m^2,i\lambda)\bigg|_{\lambda \rightarrow +\infty} \rightarrow -\frac{1}{4\pi}-\bigg(\frac{1}{4\pi}+{\cal F}^{l=k=0}_3(z^2m^2)\bigg)\frac{1}{\lambda}\nonumber \\ 
&-\bigg(\frac{1}{3\pi}+{\cal F}_1^{l=k=0}(z^2m^2)+{\cal F}_3^{l=0,k=1}(z^2m^2)+\frac{z^2m^2}{8\pi}-\frac{z^2m^2}{4}{\cal F}_2^{l=1,k=0}(z^2m^2)\bigg)\frac{1}{\lambda^2} \nonumber \\ 
&+{\cal O}\left(\frac{1}{\lambda^3}\right) \ .
\end{align}
Notice that the $-\frac{1}{3\pi \lambda^2}\equiv \frac{1}{8\pi \lambda^2}-\frac{1}{2\pi \lambda^2}+\frac{1}{24\pi \lambda^2}$ comes from 
$q_1^{l=k=0}$, $q_3^{l=0,k=1}$ and $q_2^{l=k=0}$.
Clearly, one can perform the small $z$ expansion on any of the ${\cal F}$ functions in Eq.~(\ref{eq:Ethreshold}), which leads to nothing but the large $\lambda$ expansion on top of the twist expansion. As a result, the threshold expansion for our coordinate space structure function  is consistent with the small $z$ expansion, in agreement with the recent work~\cite{Liu:2023onm}. 

\subsection{Resurgence analysis of the threshold expansion}
Unlike the twist expansion, the threshold expansion is divergent. Based on common sense, the divergence must correlate with singularities on the other side in the momentum-fraction space, such as the Reggae singularities at $\alpha=0$~\cite{Basso:2006nk}. In this subsection, we explicitly demonstrate this relation for the non-perturbative function $q_1$ in the coordinate space. 

By including all the $l=0$, $k \ge 0$ terms in the Eq.~(\ref{eq:q1thresoft}), the large $\lambda$ expansion for $q_1(\lambda)$ reads
\begin{align}
\lambda^2q_1(\lambda)=-\frac{1}{8\pi}\sum_{k,p \ge 0}\frac{(-1)^p\Gamma(2+k+p)\Gamma(2+k+2p)}{k!\Gamma(1+p)\lambda^{k+2p}}\int_{-\infty}^{\infty} dy \frac{1}{y^2+\pi^2}\rho(y)^p \ ,
\end{align}
where
\begin{align}
\rho(y)=\frac{1}{4\cosh^2 \frac{y}{2}} \le \frac{1}{4} \ .
\end{align}
From the above, it is clear that the threshold expansion grows factorially and is only a divergent asymptotic expansion. Now, one introduces the Borel transform ${\cal B}(\lambda^2q_1)(t)$
\begin{align}
{\cal B}(\lambda^2q_1)(t)=-\frac{1}{8\pi}\sum_{k,p \ge 0}\frac{(-1)^p\Gamma(2+k+p)\Gamma(2+k+2p)}{k!p!(k+2p)!}\int_{-\infty}^{\infty} dy \frac{t^{k+2p}}{y^2+\pi^2}\rho(y)^p \ .
\end{align}
Using the fact that 
\begin{align}
\bigg|\frac{\Gamma(2+k+p)\Gamma(2+k+2p)}{k!p!(k+2p)!}\bigg|\le 4^{k+p}(1+k+p)(1+k+2p) \ , 
\end{align}
the double sum for the Borel transform is absolutely convergent for $|t|<\frac{1}{4}$. 
Performing the sum first over $k$, then over $p$, one obtains the following integral representation
\begin{align}
{\cal B}(\lambda^2q_1)(t)=-\frac{1}{4\pi}\int_{0}^{\infty} \frac{dy}{y^2+\pi^2}\frac{1+t-3\rho(y)t^2}{(1-t+\rho(y)t^2)^3} \ . 
\end{align}
We now investigate its singularity structure in the positive real axis. For $0<t<1$, the $P(t,y)=1-t+\rho(y)t^2$ is always positive. Thus, the Borel transform has no singularity below $t<1$. Now, for any $0<\rho(y)\le \frac{1}{4}$, there are always two zeros $1<t_-(y)\le 2 \le t_+(y)<\infty$ of the polynomial $P(t,y)$, and as $y\rightarrow \infty$, $t_-(y)\rightarrow 1$, $t_+\rightarrow \infty$:
\begin{align}
t_{\pm }(y)=1+e^{ \pm y} \ .
\end{align}
This leads to the following analyticity structure of the Borel transform: there is a branch cut of the Borel transform along $(1,\infty)$, and the leading Borel ambiguity is of the order $e^{-\lambda}$, consistent with the ``small-$x$'' asymptotics.

We now calculate the Borel discontinuity of the integral
\begin{align}
(\Delta q_1)(\lambda)=\frac{1}{\lambda}{\rm Disc}\int_{0}^{\infty} dt e^{-t\lambda} {\cal B}(\lambda^2q_1)(t) \ ,
\end{align}
defined as the different between the $(1+i
\epsilon)t$ and the $(1-i\epsilon)t$ paths.  
To calculate the discontinuity, notice that the Borel integrals converge absolutely for $\epsilon>0$. Thus, one can exchange the $t,y$ integrals and pick up the residues at $t_{\pm}(y)$. The contribution from  $t_-(y)$ reads 
\begin{align}
(\Delta q_1)_-(\lambda)=\frac{ie^{-\lambda}}{2}\int_{0}^{\infty}\frac{dy}{y^2+\pi^2}\bigg(\frac{\left(e^y+1\right)^3}{\left(e^y-1\right)^3}-\lambda \frac{e^{-y} \left(e^y+1\right)^3}{\left(e^y-1\right)^2}\bigg)e^{-e^{-y}\lambda} \ , \\
\end{align}
and from $t_+(y)$ reads
\begin{align}
(\Delta q_2)_+(\lambda)=-\frac{ie^{-\lambda}}{2}\int_{0}^{\infty}\frac{dy}{y^2+\pi^2}\bigg(\frac{\left(e^y+1\right)^3}{\left(e^y-1\right)^3}+\lambda\frac{ \left(e^y+1\right)^3}{\left(e^y-1\right)^2}\bigg) e^{-e^{y}\lambda} \ .
\end{align}
Notice that the $y=0$ singularities cancel among the two integrals. To further simply the expressions, one changes the variable from $y$ to $\alpha$ in the following way: in the $t_-$ branch, we use $\alpha=-e^{-y}$, and in the $t_+$ branch, we use $\alpha=-e^{y}$. The two branches combine to the negative real axis $-\infty<\alpha<0$. In terms of the $\alpha$, one has
\begin{align}
(\Delta q_1)_-(\lambda)=-\frac{ie^{-\lambda}}{2}\int_{-e^{-\epsilon}}^{0}\frac{d\alpha}{\alpha}\frac{1}{\ln^2(-\alpha)+\pi^2}\bigg(\frac{(1-\alpha)^3}{(1+\alpha)^3}-\lambda \frac{(1-\alpha)^3}{(1+\alpha)^2}\bigg)e^{\alpha\lambda} \ , \\
(\Delta q_1)_+(\lambda)=-\frac{ie^{-\lambda}}{2}\int_{-\infty}^{-e^{\epsilon}}\frac{d\alpha}{\alpha}\frac{1}{\ln^2(-\alpha)+\pi^2}\bigg(\frac{(1-\alpha)^3}{(1+\alpha)^3}-\lambda \frac{(1-\alpha)^3}{(1+\alpha)^2}\bigg)e^{\alpha\lambda} \ ,
\end{align}
which combines to a ``single'' integral along $(-\infty,0)$. Notice that the $\epsilon \rightarrow 0^+$ limit of their addition is finite.

Given the above, one can now address the question: is it possible to understand the Borel discontinuity directly in terms of the explicit representation for $q_1(\lambda)$ given in Eq.~(\ref{eq:q1full})?  
For this purpose, one can use $\frac{d}{d\alpha}\frac{1}{\ln\alpha}=-\frac{1}{\alpha \ln^2\alpha}$ and partial integrates in the first integral of Eq.~(\ref{eq:q1full}) to obtain the representation
\begin{align}
q_1(\lambda)=-\frac{e^{-\lambda}}{4\pi}\int_{0}^{1}\frac{d\alpha}{\alpha}\frac{1}{\ln \alpha}\bigg(\frac{(1-\alpha)^3}{(1+\alpha)^3}-\lambda \frac{(1-\alpha)^3}{(1+\alpha)^2}\bigg)\left(e^{\alpha\lambda}-1\right)+a_1'\lambda+a_2'\ , 
\end{align}
where $a_1'$ and $a_2'$ are finite numbers related to subtraction terms for which we do not need their explicit values. It is clear from this representation, that up to subtraction terms supported at $\alpha=0$, the Borel discontinuity exactly corresponds to the branch discontinuity in the $\alpha$ space, when one extends the $\alpha$ integral to $(-\infty,0)$, namely
\begin{align}
\frac{1}{\ln (\alpha-i0^+)}-\frac{1}{\ln (\alpha+i0^+)} \rightarrow \frac{2\pi i}{\ln^2(-\alpha)+\pi^2} \ .
\end{align}
To some extent, this is expected in the following sense: the Borel transform for the $\frac{1}{\lambda}$ threshold expansion, roughly speaking, equals to the distribution in the momentum fraction space with the identification $t=1-\alpha$, up to terms that are supported away from $\alpha=1$. Therefore, the Borel singularities for the threshold expansion exactly correspond to singularities in the analytically continued $\alpha$ space distribution, also up to terms that are supported away from $\alpha=1$. In our case, the Borel discontinuity correctly anticipates the $\frac{1}{\ln \alpha}$ branch singularity at $(-\infty,0)$, and one 
can show that it equals the difference between the two $\alpha$ integrals below and above the branch cuts, including the treatment of the  $\alpha=-1$ singularity. On the other hand, the Borel discontinuity is blind to the precise form of the subtraction terms at $\alpha=0$, and this is also expected: one can add to the original distribution any terms which are Fourier transforms of distributions supported in $(0,1-\epsilon)$ without changing the algebraic threshold asymptotics. 

 \section{Coordinate space and momentum space expansions}
In this section, we investigate the issue regarding the relationship between the small-$z^2$ expansion and the large $p^2$ expansion. This issue is usually very confusing to beginners. For illustration purpose we consider the $\langle \sigma(x)\sigma(0)\rangle$ propagator at the order $\frac{1}{N}$. Throughout this section, we use the Euclidean signature for all vectors. 

We start with the representation of the Euclidean propagator
\begin{align}
\langle \sigma(p)\sigma(0)\rangle=\frac{2\pi}{N}\int_{0}^{\infty} dt \int_{c-i\infty}^{c+i\infty} \frac{ds}{2\pi i}\frac{\Gamma(1-2s)\Gamma(s+t)}{\Gamma(1-s+t)}\left(\frac{m^2}{p^2}\right)^{-s} \ .
\end{align}
Picking up poles at $s=-t-l$, one obtains the well known large $p^2$ expansion of the Euclidean propagator
\begin{align}\label{eq:expandmom}
\langle \sigma(p)\sigma(0) \rangle =\frac{2\pi}{N}\sum_{l=0}^{\infty}\frac{(-1)^l}{l!}\int_{0}^{\infty}dt\frac{\Gamma(1+2l+t)}{\Gamma(1+l+t)}\left(\frac{m^2}{p^2}\right)^{t+l} \ .
\end{align}
It has no renormalons at all and has only ``hard kernels''. We now investigate the same correlator in the coordinate space. Using the $t\rightarrow t+1$ shifting argument as in the previous sections, one has for any Schwartz-class test function $\varphi(p)$
\begin{align}
\int \frac{d^2p}{(2\pi)^2}\langle \sigma(p)\sigma(0)\rangle\varphi(p)=&\frac{2\pi}{N}\int_{1}^{\infty} dt \int_{c-i\infty}^{c+i\infty} \frac{ds}{2\pi i}\Gamma(1-2s)r(s,t)\int \frac{d^2p}{(2\pi)^2}\left(\frac{m^2}{p^2}\right)^{-s}\varphi(p) \nonumber \\
+&\frac{2\pi}{N}\int_{1}^{\infty} dt \int_{c'-i\infty}^{c'+i\infty} \frac{ds}{2\pi i}\Gamma(1-2s)\frac{1}{t^{1-2s}}\int \frac{d^2p}{(2\pi)^2}\left(\frac{m^2}{p^2}\right)^{-s}\varphi(p)  \ .
\end{align}
Notice in the second line we chose $-\frac{1}{2}<c'<0$ to guarantee the absolute convergence to support the change of order. The two terms are all bounded by $C\times{\rm sup}_{p}|(p^2+1)^{\frac{1}{2}}\varphi(p)|$, thus tempered. One now transforms it into the coordinate space. Because Schwinger functions need only to be defined outside the origin, it is sufficient to consider test functions for which $(D^\alpha\varphi)(0)\equiv 0$ for $|\alpha|\ge 0$. In this case, one has 
\begin{align}
\frac{(m^2)^{-s}}{4\pi \Gamma(-s)}\int_{0}^{\infty}d\rho\rho^{-s-2}e^{-\frac{z^2}{4\rho}}=\frac{\Gamma(1+s)}{\pi z^2\Gamma(-s)}\left(\frac{z^2m^2}{4}\right)^{-s} \ .
\end{align}
It vanishes for $s=0$. Thus for $z^2>0$ one has 
\begin{align}
\langle\sigma(x)\sigma(0)\rangle=\frac{2}{Nz^2}\int_{0}^{\infty} dt \int_{c-i\infty}^{c+i\infty} \frac{ds}{2\pi i}\frac{\Gamma(1+s)\Gamma(1-2s)\Gamma(s+t)}{\Gamma(-s)\Gamma(1-s+t)}\left(\frac{z^2m^2}{4}\right)^{-s} \ .
\end{align}
Notice we have shifted back to the original $t$ variable and the $s$-contour, which is possible again due to lacking the $s=0$ pole. This method can also be used in the main text to support the similar change of integration orders for the ${\cal E}(z^2m^2,i\lambda)$. One now shifts the contour to the left to perform the power expansion. The hard function at power $l$ reads 
\begin{align}\label{eq:hardsigmacoordinate}
{\cal H}_{\sigma\sigma}^{l}(t,\alpha(z))=\frac{2}{Nz^2}\frac{(-1)^l}{l!}\frac{\Gamma(1-l-t)}{\Gamma(t+l)}\frac{\Gamma(1+2l+2t)}{\Gamma(1+l+2t)}\left(\frac{z^2m^2}{4}\right)^{l+t} \ ,
\end{align}
while the ``soft functions'' are ($l\ge 1$)
\begin{align}\label{eq:softsigma}
{\cal S}_{\sigma\sigma}^{l}(z^2m^2,t)=\frac{2}{Nz^2}\frac{(-1)^{l-1}}{(l-1)!}\frac{\Gamma(1+2l)\Gamma(t-l)}{\Gamma(l)\Gamma(1+l+t)}\left(\frac{z^2m^2}{4}\right)^{l} \ .
\end{align}
All the $t=n$ renormalon poles cancel out as before. Moreover, the expansion is absolutely convergent again. To recover the momentum space expansion Eq.~(\ref{eq:expandmom}) at large $p^2$, one can neglect all the soft contributions (since they transform to $\nabla^{2l} \delta^2(p)$), and performs the Fourier transforms from $z$ to $p$ {\it under the Borel integrals} on the hard kernels Eq.~(\ref{eq:hardsigmacoordinate}), despite the fact that if one performs the Borel integrals first, the ``hard contributions'' at individual powers may not be Fourier-transformable. 

On the other hand, it is much harder to reproduce the purely non-perturbative terms Eq.~(\ref{eq:softsigma}) in the small-$z$ OPE by Fourier transforming the momentum space expansion Eq.~(\ref{eq:expandmom}) term by term (it is possible to reproduce all the ``non-protected ones'' such as the hard kernels or expressions with logarithms). One can try to use the ``resurgent-principle'' in the current situation: the renormalon pole of ${\cal H}_{\sigma\sigma}^{l}$ at $t=n$ must cancel with the one of ${\cal S}_{\sigma\sigma}^{l+n}$. In this way, one determines the ${\cal S}_{\sigma\sigma}^{l}$ up to analytic functions in neighborhoods of the positive real axis. Notice that outside the $\frac{1}{N}$ expansion, it is indeed difficult for operators to be completely free from perturbative coefficients in non-SUSY theories, but the OPE coefficients at any power, on top of the Borel integrals, may also have log-free terms. Clearly, these terms can not be reproduced by Fourier transforming term by term in the large $p^2$ expansion.

One can summarize the following lessons from the above example for practical perturbative calculations: it is safer to Fourier transform from the coordinate space to the momentum space than the opposite direction. It is correct to Fourier transform at the level of fixed-order PT series with respect to $\alpha(\mu)$ and then RGE resum. It is also correct to Fourier transform under Borel representations. But {\it it is usually questionable to perform the Fourier or other integral transforms at the level of RGE re-summed expressions, unless they can be converted into Borel representations.} Moreover, renormalons generated during such integral transformations, based on truncated Borel integrands, can be misleading: $\frac{1}{\Gamma(1-t)}$ when expanded near $t=0$ and then Fourier transformed may lead to fake renormalons at $t=1$. 

The correctness of Fourier transforming under the Borel representations can be argued in the following sense: the coordinate-space correlators are analytic in the ${\rm Re}(z^2)>0$ region and decay exponentially fast at infinity. Thus, one expects the existence of Mellin representations of the form
\begin{align}
{\cal W}(z^2)=\frac{1}{|z|^{2\Delta}}\int_{c-i\infty}^{c+i\infty} \frac{ds}{2\pi i}{\cal M}(s)\left(\frac{z^2m^2}{4}\right)^{-s} \ .
\end{align}
Here, since we have factorized out the most singular part of the two-point function, the $c$ can be chosen to be any positive number. In particular, ${\cal M}(s)$ should be analytic in the right-half plane and has branch cuts and poles (either isolated or on top of the branch cuts) only in the negative real axis. Moreover, the representation should hold at least in the ``below-threshold region'' ${\rm Re}(z^2)>0$, 
implying that ${\cal M}(s)$ should decay sufficiently fast when ${\rm Im}(s) \rightarrow \pm \infty$, which we assume to hold for ${\rm Re}(s)<0$ as well. Let's consider the case $D>2\Delta>0$ where $D$ is the space-time dimension. In this case, the Schwinger function is tempered without subtractions. Then, one can  chose $0<{\rm Re }(s)<\frac{D}{2}-\Delta$ to Fourier transform under the Mellin representation to obtain
\begin{align}
{\cal W}(p^2)=\frac{(4\pi)^{\frac{D}{2}}}{2^{2\Delta}|p|^{D-2\Delta}}\int_{c-i\infty}^{c+i\infty} \frac{ds}{2\pi i}\frac{\Gamma(\frac{D}{2}-\Delta-s){\cal M}(s)}{\Gamma(\Delta+s)}\left(\frac{m^2}{p^2}\right)^{-s} \ ,
\end{align}
for $p^2>0$. One now shifts the contour to the left to perform the large $p^2$ or small $z^2$ expansions. The point is that no new singularities are generated during the Fourier transform from the coordinate space to the momentum space. Thus, when picking up the branch discontinuities in the negative real axis to form the asymptotic expansions, the expressions for the Borel integrals in the two spaces look exactly like Fourier transforms of each other for the Borel integrands.

On the other hand, for the ``isolated'' singularities (which can be on top of branch cuts), this also holds with one exception:  the factor $\frac{1}{\Gamma(\Delta+s)}$ contains extra zeros 
at $s=-\Delta-n$, $n\in Z_{\ge 0}$, if the ``Mellin amplitude'' ${\cal M}(s)$ also has simple poles (either isolate or on top of branch cuts) at these locations, then in the coordinate space they contribute to terms proportional to $\left(\frac{z^2m^2}{4}\right)^n$ without extra logarithms, while in the momentum space, they become ``delta-ditributions'' at $p=0$ and disappear in the large $p^2$ expansion. Notice that if $2\Delta=D$ and ${\cal M}(s)$ is bounded near $s=0$ with ${\rm Disc}{\cal M}(0^-)=0$, then the coordinate space correlator is also tempered (through inverse-log square) and arguments above hold as well, which exactly corresponds to the case of $\langle \sigma(x)\sigma(0)\rangle$ discussed above. Moreover, from the above, one can also see that in the $t \rightarrow -\infty$ limit, the Mellin-amplitude along the path $s=(1+i\epsilon)t$ in the momentum-space is enhanced by two factors of gamma functions. If one believes that the large $p^2$ expansion does not grow extremely fast to the extent that does not even allow convergent Borel sums near the origin, then ${\cal M}((1+i\epsilon)t)$ should decay when $t \rightarrow -\infty$ at least at the inverse factorial speed, strongly suggesting that the small $z^2$ expansion should converge absolutely for any $z^2>0$. 

To illustrate the arguments above, let's consider an example with ``infinitely-many bound states'', the 2D QCD in the planar limit~\cite{tHooft:1974pnl}. This type of example has been used in the literature before~\cite{Boito:2017cnp} to investigate possible ``non-universalities'' of the large $p^2$ expansions across space-like and time-like regions. The ``polarization function'' defined through the vacuum expectation of two vector currents in the theory can be represented as~\cite{Einhorn:1976uz} (in mass unit with $\frac{g^2N}{\pi}=1$)
\begin{align}
\frac{\pi p^2}{N}\Pi_V(p^2)-1 \equiv G(p^2)=-\sum_{n=1}^{\infty}\left(\frac{2m}{m_n}\right)^2\frac{g_n^2}{p^2+m_n^2} \ ,
\end{align}
where $m$ is the quark mass, $m_n$ are the bound state masses and $g_n$ are the coupling constants between the vector current and the bound states. We would like to understand the short distance or high $p^2$ asymptotics of the above correlator. To proceed, we transform to the coordinate space and introduce the Mellin representations for the Bessel functions. This leads to
\begin{align}
&G(z^2)=\frac{m^2}{\pi}\int_{1-i\infty}^{1+i\infty} \frac{ds}{2\pi i}\Gamma^2(s) Q(s) \left(\frac{z^2}{4}\right)^{-s} \ , \\ 
&Q(s)=-\sum_{n=1}^{\infty}\frac{g_n^2}{m_n^{2s+2}} \ .
\end{align}
Notice that `` Dirichlet sums'' like the one for $Q(s)$ here also appeared  in~\cite{deRafael:2010ac} to represent the polarization function in the momentum space.  It is known~\cite{Einhorn:1976uz,Fateev:2009jf} that in the $n \rightarrow \infty$ limit, $m_n^2 \rightarrow \pi^2 (n+\frac{3}{4})+{\cal O}\left(\frac{1}{n^2}\right)$, $g_n^2 \rightarrow \frac{\pi^2}{2}$. As a result, $Q(s)$ defines an analytic function in $s$ in the right-half plane and develops a singularity at $s=0$. Assuming that the subleading asymptotics for $g_n$ and $m_n$ are not extremely pathological, $Q(s)$ can be further continued to the left as a meromorphic function. In particular, the leading large $n$ asymptotics leads to
\begin{align}\label{eq:Qasym}
Q^{\rm asym}(s)=-\frac{1}{2\pi^{2s}}\zeta \left(s+1,\frac{7}{4}\right) \ .
\end{align}
It has a simple pole at $s=0$, leading to the famous $\frac{1}{p^4}\ln p^2$ subleading contribution to $\Pi_V(p^2)$, but otherwise is analytic. 

What is the high power behavior in the small $z^2$ or large $p^2$ expansions controlled by $Q^{\rm asym}(s)$? In the small $z^2$ expansion, since the $\zeta$ function only introduces one factor of factorial growth, while the $\Gamma^2(s)$ is suppressed by two factors, {\it the convergence radius is still infinite}. In the momentum space, on the other hand, the $l$-th power reads
\begin{align}
G^{(l)}(p^2)=\frac{2m^2}{p^2}\left(\frac{\pi^2}{p^2}\right)^l\frac{(-1)^lB_l\left(\frac{7}{4}\right) }{ l} \ , l\ge 1 \ , 
\end{align}
which grows factorially, but is still Borel-summable as the singularities of the Borel transform are in the imaginary axis~\cite{Boito:2017cnp}. Notice that here we only included the leading large $n$ asymptotics for the spectral data. Using the exact profile may not change dramatically the overall factorial speed, but may change the oscillations in the subleading terms that could move the singularities of the Borel transform from the imaginary to the real axis. To determine the precise fate of the large $p^2$ expansion in the planar 2D QCD, therefore, requires more detailed investigations.

We should also mention that given the results in this section, one can ``reproduce'' the momentum space expansion for the fermion self-energy recently performed in Ref.~\cite{Marino:2024uco}. The Euclidean fermion propagator reads ($\vec{\gamma}=(\gamma^0,i\gamma^1)$)
\begin{align}
S_F(z,m)=\frac{1}{2\pi}(\slashed\partial+m)K_0(mz)=\frac{1}{4\pi} (\slashed\partial+m)\int_{1-i\infty}^{1+i\infty} \frac{ds}{2\pi i}\Gamma^2(s) \left(\frac{m^2z^2}{4}\right)^{-s}\ .
\end{align}
The double-pole leads to the $\ln z^2m^2$ in the small-$z$ expansion:
\begin{align}
K_0(mz)=\frac{1}{2}\sum_{k=0}^{\infty}\frac{1}{k!^2}\left(\frac{z^2m^2}{4}\right)^k \bigg(-\ln \frac{m^2z^2e^{2\gamma_E}}{4}+2H_{k}\bigg) \ .
\end{align}
Again, the convergence radius is infinite. Multiplying the above with the expansion of the $\langle \sigma(x)\sigma(0) \rangle$ leads to the small $z$ expansion of the fermion self-energy. Performing the Fourier transforms at the level of Borel integrands, one obtains the large $p^2$ expansion of the self-energy, up to renormalization-scheme ambiguities for the  singularities at $t=0$.

\section{Summary and outlook}
To summarize, in this work, we have performed the exact twist and threshold expansions for a specific structure-function in the Bjorken and the threshold limits in the 2D large-$N$ Gross-Neveu model. The hard and ``collinear'' functions in the twist expansion are generated naturally as Borel integrals, with the cancellation pattern of the $t=n$ singularities between hard and non-perturbative functions manifest. Explicit forms for the non-perturbative functions are also provided,  which are simplified in detail at the leading power. We also prove that the twist expansion is absolutely convergent for any $z^2>0$. In the threshold limit, the contributions for an arbitrary $z^2$ at the first three threshold powers are also calculated, and the resurgence relation between the threshold and the ``Regge'' asymptotics is explicitly verified for a specific non-perturbative scaling function. There is no conflict between the threshold and the twist expansions. We also discussed the relationship between the momentum space and the coordinate space expansions.

Before ending this paper, we would like to make some comparisons with the Ref.~\cite{Balog:2004mj} and Ref.~\cite{Balog:2004yg}. In these two references, the authors also investigated the Bjorken limit of structure function type correlators in 2D QFTs, in particular, at the level of the $\frac{1}{N}$ expansion in the $O(N)$ non-linear sigma models. On the other hand, the formulation of the Bjorken limit in these references is the traditional version in the momentum space with $q^2$ and $x$. As a result, for correlators like the one in Eq.~(\ref{eq:def}), at the ``one bubble-chain'' diagram level, one is essentially probing the discontinuity of the $\sigma$ propagator along the cut $(4m^2,\infty)$, which contains no non-perturbative scaling functions like the $q_1^{(l)}(\lambda,\mu)$ in Eq.~(\ref{eq:q1lmu}) at all. To see the coexistence of both the perturbative and non-perturbative functions in the momentum space formulation, one must calculate diagrams with two $\sigma$ propagators for the correlator in Eq.~(\ref{eq:def}). To some extent, the coordinate space correlator in Eq.~(\ref{eq:def}) is almost the simplest one that is still complicated enough to exhibit many essential features of the Bjorken limit at the level of one bubble-chain diagrams, in particular, the existence of PDF-like non-perturbative scaling functions even at the leading power. 

There are several possible ways to consolidate and extend the results of this work further. It is possible and quite essential to verify that all the ``hard kernels'' obtained here can indeed be reproduced through the massless OPE calculations (in the same spirit as what has been performed recently for the fermion self-energy in~\cite{Marino:2024uco}). It is also possible and perhaps more important to show that all the obtained non-perturbative scaling functions here can indeed be reproduced based on naive operator definitions as light-front correlators. Due to the large number of operators at any given twist and the presence of UV divergences for these operators, this task is also not entirely trivial. One should also notice that despite the large number of operators, the final results at each power are still simple and compact. This implies that there might be specific ways to reorganize or recombine these operators to simplify the results, at least to the particular order we reached. Finally, one should mention that the PDF-like non-perturbative scaling functions are usually regarded as being universal in the sense that the same distribution should remain the same in various possible formulations of the same Bjorken limit. In particular, at least at the leading power,  it should be possible to verify that the obtained non-perturbative functions in this work can also be reproduced in the traditional momentum space formulation of the Bjorken limit.

\acknowledgments
The author thanks Dr. Ramon Miravitllas for notification of the important reference~\cite{Beneke:1998eq}. The author thanks Yushan Su for discussions and double-checking of most of the equations before Sec.~\ref{sec:threshold} of the previous version. Y. L. is supported by the Priority Research Area SciMat and DigiWorlds under the program Excellence Initiative - Research University at the Jagiellonian University in Krak\'{o}w. 

\appendix

\bibliographystyle{apsrev4-1} 
\bibliography{bibliography}

\end{document}